%% file: main.tex
\documentclass[sigconf,nonacm]{acmart}
\AtBeginDocument{%
  }

\usepackage{amsmath}
\usepackage{algorithmic}
\usepackage{booktabs}
\usepackage{url}
\usepackage{amsfonts}
\usepackage{graphicx}
\usepackage{multirow}
\usepackage{hyperref}
\usepackage{subfigure}

\begin{document}

\title{Re-pseudonymization Strategies for Smart Meter Data
Are Not Robust to Deep Learning Profiling Attacks}

\author{Ana-Maria Cretu}
\authornote{Equal contribution.}
\authornote{Most of the work was done while the author was at Imperial College London.}
\affiliation{
  \institution{EPFL}
  \streetaddress{P.O. Box 1212}
  \city{Lausanne}
  \country{Switzerland}
}
\email{ana-maria.cretu@epfl.ch}

\author{Miruna Rusu}
\authornotemark[1]
\affiliation{
  \institution{Imperial College London}
  \city{London}
  \country{United Kingdom}}
\email{miruna.rusu17@alumni.imperial.ac.uk}

\author{Yves-Alexandre de Montjoye}
\affiliation{
  \institution{Imperial College London}
  \city{London}
  \country{United Kingdom}}
\email{deMontjoye@imperial.ac.uk}

\begin{abstract}
Smart meters, devices measuring the electricity and gas consumption of a household, are currently being deployed at a fast rate throughout the world. 
The data they collect are extremely useful, including in the fight against climate change.
However, these data and the information that can be inferred from them are highly sensitive. 
Re-pseudonymization, i.e., the frequent replacement of random identifiers over time, is widely used to share smart meter data while mitigating the risk of re-identification.
We here show how, in spite of re-pseudonymization, households' consumption records can be pieced together with high accuracy in large-scale datasets.
More specifically, we propose the first deep learning-based profiling attack against re-pseudonymized smart meter data.
Our attack combines neural network embeddings, which are used to extract features from weekly consumption records and are tailored to the smart meter identification task, with a nearest neighbor classifier.
We evaluate six neural networks architectures as the embedding model.
Our results suggest that the Transformer and CNN-LSTM architectures vastly outperform previous methods as well as other architectures, successfully identifying the correct household 73.4\% of the time among 5139 households based on electricity and gas consumption records (54.5\% for electricity only).
We further show that the features extracted by the embedding model maintain their effectiveness when transferred to a set of users disjoint from the one used to train the model.
Finally, we extensively evaluate the robustness of our results. In particular, we show that the accuracy of the attack only slowly decreases with the size of the dataset, how less frequent re-pseudonymization will further increase accuracy, and how an attacker can evaluate the likelihood of a match to be correct. 
Taken together, our results strongly suggest that even frequent re-pseudonymization strategies can be reversed, strongly limiting their ability to prevent re-identification in practice.\footnote{This is an extended version, including the Appendix, of our paper which will appear in the proceedings of the ACM CODASPY 2024 conference.}
\end{abstract}

\maketitle
\pagestyle{plain}

\input{sections/introduction.tex}

\input{sections/attack_model}

\input{sections/related_work}

\input{sections/attack_methodology}

\input{sections/experimental_setup}

\input{sections/results}

\input{sections/discussion}

\input{sections/conclusion}

\textbf{Acknowledgements}
The authors thank Julien Baur, David Ferguson and Nick Dillon as well as the EDF UK R\&D Digital Innovation team for providing access to the data and for the discussions.
This work was supported by the PETRAS National Centre of Excellence for IoT Systems Cybersecurity, funded by the UK EPSRC under grant number EP/S035362/1.

\bibliographystyle{plain}
\bibliography{bibliography}

\appendix
\input{sections/appendix}

\end{document}

%% file: sections/introduction.tex
\section{Introduction}
\label{sec:introduction}

Smart meters, devices measuring the utility (electricity, gas, water, etc.) consumption of a household, are currently being deployed at a fast rate throughout the world~\cite{iotanalyticsdeployment}. 
For instance, the United Kingdom (UK) had installed almost 30M smart meters by June 2022 ~\cite{uk30Msmartmeters}, while the European Union (EU)'s Smart Grid Task Force aims to have smart meters installed in 92\% of homes by 2030~\cite{alaton2020benchmarking}. 

Smart meters are believed to be a key element in enabling more efficient management of renewable resources for electricity, but also gas or water. They provide valuable data that allow advanced load planning and optimization within the smart grid~\cite{euhorizon2020}. Additionally, they offer more flexibility on the demand side through real-time updates on consumption and prices and foster the integration of renewables and helps decarbonize the energy market~\cite{govsmartsystemsandflexibility, ofgemupgradingourenergysystem}. Finally, smart meters help people better manage their consumption~\cite{citizensadvicesmartmeters}, something likely to further drive adoption given the increase with energy prices rising across Europe~\cite{eupricerise,ukpricerise}.

While very useful, consumption data collected by smart meters can also be very sensitive. 
Both the Art 29 Working Party (predecessor of the European Data Protection Board) and the European Data Protection Supervisor emphasized the intrusiveness of smart meters in people’s life in their respective guidance regarding smart meter systems~\cite{wp2011,edps}. 
Typically collected at high-resolution time intervals, e.g., every half an hour~\cite{topping2021}, smart meter readings contain fine-grained information about life in the house. It has been shown, for instance, that household daily patterns such as wake-up and bedtimes~\cite{lisovich2008privacy} can be inferred from electricity consumption records. Other information~\cite{wang2018deep} such as occupancy status~\cite{razavi2019occupancy}, the number of inhabitants, their relationship, unemployment~\cite{montanez2020machine} and socio-economic status~\cite{beckel2013automatic,beckel2014revealing,fahim2019analyzing}, specific appliances being used~\cite{rottondi2019optimisation}, and channels being watched~\cite{greveler2012multimedia},  were also shown to be  predictable from smart meter readings under certain conditions. 

To allow smart meter data to be used without compromising privacy, individual records are typically pseudonymized before they are shared~\cite{energyrevprivacy}. This means that personal identifiers are removed and every user (household) is instead assigned a consistent random number (pseudonym). For instance, Ms. Jane Doe would be assigned the number 074375 and all her records would be identified by this number. 

Pseudonymization has, however, been shown to be ineffective on its own to prevent re-identification by attackers. Re-identification attacks have, for instance, proposed to uniquely match a target's (known) aggregate consumption over one or several months~\cite{cleemput2018pseudonymization} to the pseudonymized data. 
Other attacks, leveraging the high unicity of smart meter data, have proposed to re-identify user through records of their consumption at specific times~\cite{voyez2022unique}.
For instance, if the attacker knows that Ms. Doe consumed 883kWh of electricity in December 2023, and there is only one pseudonymous record that matches this consumption value, the attacker can now retrieve her fine grained, e.g., every 5 min to an hour, consumption data. They can then learn new information by analyzing the target's consumption patterns~\cite{lisovich2008privacy, razavi2019occupancy, montanez2020machine, beckel2013automatic,beckel2014revealing,fahim2019analyzing, rottondi2019optimisation, greveler2012multimedia}.

To prevent such \textit{matching} attacks, and in an attempt to make smart meter data anonymous, the data is typically re-pseudonymized and/or generalized before being shared~\cite{cleemput2018pseudonymization,ofgembaringa}. 
Generalization means that consumption is recorded less frequently, or that frequent readings are aggregated before they are released, e.g., at an hourly instead of half-hourly level.
Re-pseudonymization means that the unique pseudonym assigned to a user is changed over time, typically on a weekly or bi-weekly basis. In our example, while in the first week of December Ms. Doe would be user 074375, in the second week she would be user 076468. Re-pseudonymization aims to mitigate the risk of re-identification posed by matching attacks by both a) limiting the amount of information recovered when an attack succeeds and by b) limiting the period of time over which an adversary can collect auxiliary information for a matching attack.

For instance, assume that an attacker has access to smart meter data from many households in a month, including Ms. Doe's, and that records are re-pseudonymized on a weekly basis.
To perform a unicity attack, the attacker would have to acquire sufficient auxiliary data to uniquely identify Ms. Doe within a given week. If successful, this would give them access to one week of Ms. Doe's data. The attacker would then need to acquire enough data again for a second week and so on. While weekly re-pseudonymization does not prevent unicity-based re-identification attacks, it makes them materially more difficult especially if one aims to obtain more than one week of the target's data. Furthermore, weekly re-pseudonymization entirely prevents what we argue is the most likely re-identification attack: using monthly consumption data, a fairly widely available piece of information, to uniquely identify a target user. 
Taken together, re-pseudonymization and generalization have been argued to be sufficient to anonymize data, making it out of scope of the GDPR. Importantly however, while pseudonymization does not impact downstream uses of the data, e.g., load forecasting~\cite{shi2017deep} or outage management~\cite{jiang2015outage}, re-pseudony- mization and generalization can reduce the general utility of the data, e.g., by preventing longitudinal studies.

Profiling attacks, inspired by facial recognition techniques, have however been developed for a range of behavioral data~\cite{mcilroy2021detecting,crectu2022interaction,tournier2022expanding}. In our example, an attacker could use a profiling attack to infer that user 074375 in the first week of December is very likely the same person as user 076468 in the second week of December. If the profiling attack succeeds for the three pairs of consecutive weeks in a month, the attacker obtains a monthly record for every pseudonym.
The attacker can then run a matching attack using monthly consumption as the auxiliary information.

Profiling attacks for smart meter data have however, so far, relied on standard classification techniques, e.g., support vector machines~\cite{jawurek2011smart} or similarity search using hand-engineered features~\cite{buchmann2013re,tudor2015study,tudor2018influence}. These attacks are, we argue, not sufficient to ``reverse'' the weekly re-pseudonymization used in practice. Indeed, first, the best method so far would only have a 0.3\% chances of successfully linking back together 4 weeks of data of a person, much below the 5\% risk threshold used, e.g., by the United Nations Office for the Coordination of Humanitarian Affairs' (UN OCHA) data exchange platform HDX~\cite{ocha2019}. Second, previous attacks assume a very strong attacker with access to significant auxiliary data about the target, typically at least one month of smart meter data~\cite{jawurek2011smart,faisal2015quantity}, which we argue to be unrealistic and, very often, to defeat the purpose of re-identification in the first place as multiple weeks of data are already available to the attacker. 

\textbf{Contribution.} We address both of these limitations and propose what is, to our knowledge, the first deep learning-based profiling attack against re-pseudonymized smart meter data. 
More specifically, we focus on the problem of piecing together weekly re-pseudonymized records of a large set of households from two weeks $\mathcal{T}_1$ and $\mathcal{T}_2$.
Given a target household's smart meter record in $\mathcal{T}_2$, the goal is to identify the correct record (and pseudonym) in $\mathcal{T}_1$ corresponding to that household.
Our attack uses neural network embeddings tailored to the smart meter domain to extract a set of features from a weekly consumption record.
We train the embeddings specifically for the task of identifying users across time, using the triplet loss~\cite{schroff2015facenet,crectu2022interaction}.
We then compute the Euclidean distances between the target household (in $\mathcal{T}_2$) and all the candidate households (in $\mathcal{T}_1$), retrieving the closest candidate as the match. 
We evaluate 6 neural network architectures as the embedding model in two different attack scenarios.

We first compare the accuracy of our attack to existing work. Our first scenario therefore assumes an informed attacker having an auxiliary dataset of smart meter records from the same households as in $\mathcal{T}_1$, but collected in a disjoint time period $\mathcal{T}_{\text{aux}}$ spanning several weeks.
We train the embedding model on this auxiliary dataset and use it to run the attack against the re-pseudonymized smart meter records of the same users from periods $\mathcal{T}_1$ and $\mathcal{T}_2$, as described above.
We show our method to strongly outperform previous work, successfully identifying users 54.5\% of the time in a dataset of 5139 electricity users. 
A fair comparison with previous methods shows the best previous work to only reach 14.7\% accuracy. 

To illustrate the relevance of our results, we extrapolate the 54.5\% accuracy of matching consecutive weeks to $0.545^3=0.161=$16.1\% chance of correctly piece together one month (i.e., three pairs of consecutive weeks) of an individual's data, compared to 0.3\% for the best previous method. Both numbers give us a direct measure of the risks of re-identification as Cleemput et al.~\cite{cleemput2018pseudonymization} showed matching attacks using monthly aggregate consumption to be almost 100\% accurate when no perturbation is applied to the data.
Taking the UN OCHA's, HDX, 5\% risk as the threshold, one could have wrongly concluded that weekly re-pseudonymization successfully thwarts re-identification attacks (profiling followed by matching). Our state-of-the-art method shows this to be incorrect.

Second, we show how our state-of-the-art method alleviates the (very strong) requirement for the attacker to have access to an auxiliary dataset with past behavior from the target user.
In our second scenario, we assume that the auxiliary dataset is collected from a different set of users than the one the attacker is trying to piece together.
For instance, the attacker might gain access to a publicly available dataset~\cite{commission2012cer}, to one of the many datasets used in previous smart meter studies, e.g.,~\cite{alhussein2020hybrid}, or could be one of many subcontractors with access to such data.
This allows the attack to be performed even in frequent re-pseudonymization scenarios, e.g., weekly, with no other information about users. 
We show that the features learnt by the embedding model transfer to new households at almost no cost in accuracy, as they successfully identify households unseen by the model during training 52.2\% of the time. 
This demonstrates that only one week of data from a target household is sufficient to identify them, strongly questioning, contrary to previous results, the protection offered by re-pseudonymization strategies.

We further validate our results by conducting the largest identifiability experiment to date on smart meter data and show the accuracy of our attack to slowly decrease with the population size (number of candidate households). The attack indeed remains effective even on 67k users, reaching an accuracy of 29.2\%.  

Finally, we show that more information about consumption would further increase the risk. Assuming the attacker has access to gas consumption indeed improves the accuracy of the attack by 18 p.p. 
We also show that reducing the frequency of re-pseudonymization strongly improves the accuracy of the attack.
    
Taken together, our results strongly suggest that smart meter data is highly identifiable and re-identifiable even when re-pseudonymized weekly, strongly challenging the privacy-utility benefits achieved by re-pseudonymization strategies.

%% file: sections/attack_model.tex
\section{Background}
\label{sec:background}

We first introduce the concepts of smart meter record describing the data collected by smart meters and re-pseudonymization, the main technique used to anonymize smart meter records.
Then, we describe the profiling threat model that is used to evaluate the risk of re-identification in re-pseudonymized datasets.

\subsection{Smart meter record}
A smart meter is a device recording information about the energy consumption of an individual property on a regular basis, e.g., every 30 minutes.
Let $\mathcal{P}$ be a set of  \textit{individual properties} with a smart meter installed. 
We refer to the set of inhabitants of an individual property $p \in \mathcal{P}$ 
as the \textit{user}. 
The smart meter transmits the consumption data of one or more utilities (e.g., electricity, gas) recorded at different timestamps separated by the \textit{time granularity} $\Delta t$, e.g., 30 minutes. 
The data recorded about a utility are equal to the amount that was consumed within the last time interval. 
As the smart meter may record multiple utilities, the data can be formally described as a multivariate time series. 
For each timestamp $t$ and individual property $p \in \mathcal{P}$, the consumption values for $F$ utilities are recorded as a feature vector $r^p_t=(r^p_{t,1}, \ldots, r^p_{t,F}) \in \mathbb{R}^F$ (e.g., $F=2$ for electricity and gas and $F=1$ for electricity alone). 
The smart meter record of an individual property $p$ over a time period $\mathcal{T}=[t, t')$ having $t' = t + T \times \Delta t$ is the timeseries $R^p_{\mathcal{T}} \in \mathbb{R}^{T \times F}$ of its consumption (starting at $t$): $R^p_{\mathcal{T}} = ( r^p_{t} \;,\; r^p_{t+\Delta t} \;,\; \ldots \;,\; r^p_{t+ (T-1)\Delta t} )$.

\subsection{Re-pseudonymization} 
We consider a setting where smart meter records are shared in plaintext in order to enable the end user, e.g., an analyst, to perform data analytics.
As smart meter records are very sensitive, in order to limit the risk of re-identification, we assume that they are frequently re-pseudonymized, e.g., weekly, before they are shared.
For a smart meter record $R^p_{\mathcal{T}}$, this means that, first, the time period $\mathcal{T}$ over which the data is collected is split into disjoint sub-periods, e.g., by weeks, $\mathcal{T}=\mathcal{T}_1 \cup \ldots \cup \mathcal{T}_T$.
Second, the direct identifier of the user is removed from the dataset and replaced with $T$ identifiers $h_1(p), h_2(p), \ldots, h_T(p)$, e.g., random hashes, one for every sub-period.
These identifiers are both different from one another: $h_i(p) \neq h_j(p), \forall i \neq j$ and from the original one $h_i(p) \neq p, \forall i$.

Given a subset of individual properties $P \subset \mathcal{P}$, a time period $\mathcal{T}$, and a hashing function $h$, we denote by \textit{dataset} $D^{P, h}_{\mathcal{T}}$ the set of smart meter records of users in $P$ collected over time period $\mathcal{T}$: $D^{P,h}_{\mathcal{T}} = \{ (R^{p}_{\mathcal{T}}, h(p)): p \in P \}$, where each smart meter consumption record is identified by a unique pseudonym.

\subsection{Profiling threat model}

Profiling attacks~\cite{buchmann2013re,jawurek2011smart,tudor2013analysis,tudor2018influence} typically assume that a malicious individual, the attacker, gains access to a dataset $D_{\mathcal{T}_1}^{P_1,h_1} \cup D_{\mathcal{T}_2}^{P_2,h_2}$ consisting of the smart meter records of users $P_1$ and $P_2$ collected over two disjoint time periods $\mathcal{T}_1$ and $\mathcal{T}_2$.
The two sets of users are assumed to have at least a few users in common $P_1 \cap P_2 = P_0 \neq \emptyset$.
Importantly, the smart meter records are re-pseudonymized between the two time periods such that for every common user, their pseudonyms in the two datasets are different, i.e., $\forall p \in P_0: h_1(p) \neq h_2(p)$.
This can be achieved by hashing the concatenation of the direct identifier of the user, an identifier of the time period, and a random salt or using another method such as encryption.
The attacker aims to reverse the re-pseudonymization by retrieving, for a subset of pseudonyms of the common users between the two datasets, $Y \subset \{ h_2(p), p \in P_0\}$, their correct matches in the other dataset. 
More specifically, for every $y \in Y$, the goal is to find the unique pseudonym $y'$ in the other dataset such that $\exists p \in P_0, y=h_2(p)$ and $y'=h_1(p)$.

As directly reverting the pseudonyms is infeasible (if re-pseudo- nymization is done properly), the attacker aims to exploit unique recurring patterns in the target user's utility consumption. 
The underlying hypothesis of the profiling attack is that the consumption patterns of a household are both (1) unique enough among other households and (2) regular enough across time to enable correct linkage.
From a legal perspective, a successful linkage could, according to current guidelines of the EU Article 29 Working Party~\cite{wp2014}, constitute a breach of GDPR-anonymization. 
Furthermore, successful linkage would reverse the re-pseudonymization, putting users again at risk of a re-identification matching attack. Here, an attacker would first perform a profiling attack to piece the records back together, giving them a person's records over the extended time period $\mathcal{T}_1 \cup \mathcal{T}_2$. 
Then, the attacker would then perform a matching attack to re-identify a user. Numerous auxiliary information and techniques have been used to perform matching attacks including monthly aggregate consumption from an invoice~\cite{cleemput2018pseudonymization} or a few smart meter readings~\cite{voyez2022unique}.
Re-identification gives the attacker access to the smart meter records of the user, allowing him or her to infer sensitive private information about the inhabitants, e.g., household occupancy~\cite{razavi2019occupancy}, wake-up and bed times~\cite{lisovich2008privacy}, when people are at home, if they are unemployed~\cite{montanez2020machine}, and their socio-economic status~\cite{beckel2013automatic,beckel2014revealing,fahim2019analyzing}.

\begin{figure}[t!]
\centering
\includegraphics[width=\linewidth]{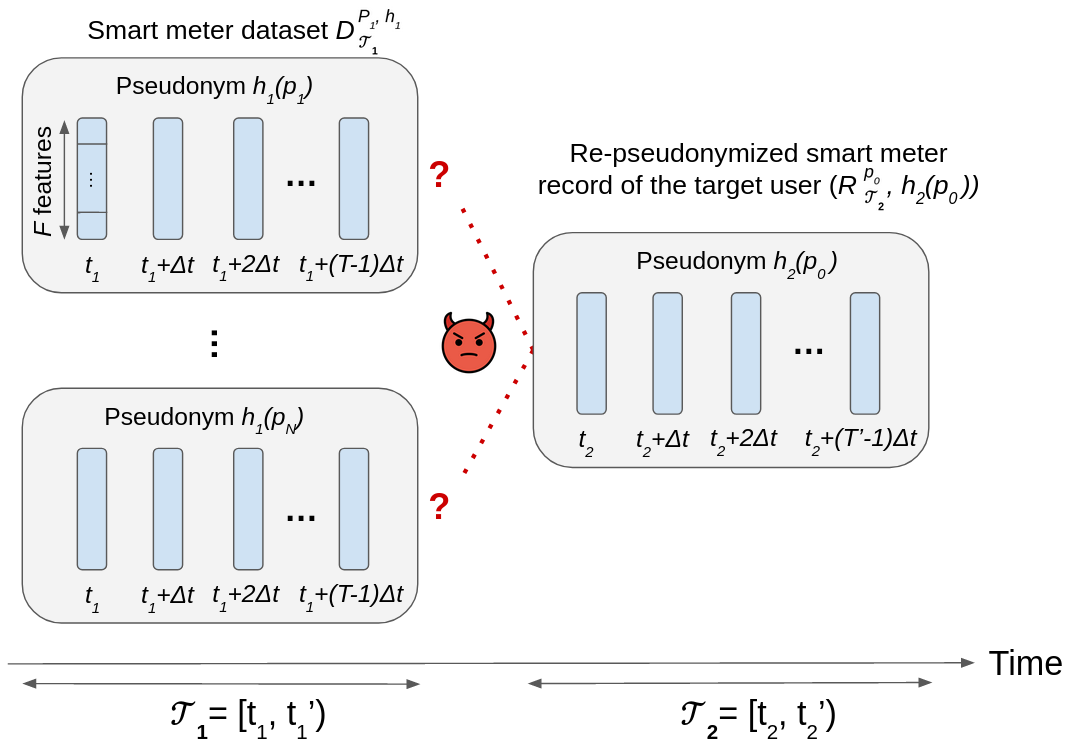}
\caption{Profiling threat model. The attacker aims to reverse the re-pseudonymization by piecing together the target user's smart meter records from disjoint time periods $\mathcal{T}_1$ and $\mathcal{T}_2$.
}
\label{figure:attack_model}
\end{figure}

Fig. \ref{figure:attack_model} illustrates the profiling threat model, instantiated on one individual property $p_0 \in P_0$, referred to as the \textit{target user}. 
The goal of the attacker is to retrieve the record $R^{p}_{\mathcal{T}_1}$ such that $p=p_0$ using as \textit{auxiliary information} the target user's record in a disjoint time period, $R^{p_0}_{\mathcal{T}_2}$. 
If successful, the attacker is able to link together the consumption records from the two time periods, $(R^{p_0}_{\mathcal{T}_1}, h_1(p_0))$ and $(R^{p_0}_{\mathcal{T}_2}, h_2(p_0))$. 
We will henceforth call $P_1$ the \textit{reference user set}.

%% file: sections/related_work.tex
\section{Related work}
\label{sec:related_work}

\textbf{Profiling attacks against smart meter data.} 
Jawurek et al.~\cite{jawurek2011smart} proposed the first profiling attack linking together smart meter consumption records from different time periods. 
The method used support vector machines to classify smart meter readings based on the histogram of daily consumption. 
While their attack achieved high accuracy, it was only evaluated on a dataset of 53 reference users.
Buchmann et al.~\cite{buchmann2013re} proposed another approach relying on 12 hand-engineered features and similarity-based matching. 
The attack was here evaluated on a dataset of 36 reference users and two consecutive weeks. 
Tudor et al.~\cite{tudor2015study} evaluated the scalability of this approach when applied to larger user sets (of up to 400 users, and later 2000 users ~\cite{tudor2018influence}), showing the attack accuracy to decrease as a function of the reference user set size. 
Their work simplified Buchmann's method by removing the parameters and reducing the number of features to 5. 
Bicego et al.~\cite{bicego2014behavioural} extracted a different set of hand-engineered features and used distance-based matching as well as Hidden Markov models. 
Faisal et al.~\cite{faisal2015quantity} used a decision tree classifier and applied the attack to two small-scale user sets, of sizes 26 and 100, respectively.

These methods have only been shown to be effective against small-scale datasets, making it unclear how the risk of re-identifica- tion scales to large population sizes.
Moreover, these methods very likely do not provide a tight estimate of the risk. 
This is because they rely on hand-engineered features and standard machine learning classification models that have been outperformed by deep neural networks and, in particular, embeddings on numerous classification tasks, including identification in other data domains~\cite{crectu2022interaction,fabien2020bertaa,mcilroy2021detecting}.
We here evaluate for the first time the effectiveness of state-of-the-art neural network embeddings of smart meter data for identification.
We strongly outperform previous work and show profiling attacks to be a risk even when little information is available, as is typically the case when re-pseudonymization strategies are used. 

Shateri et al.~\cite{shateri2019deep,shateri2021learning} explored the use of privacy-preserving feature representations as a substitute for sharing raw smart meter data.
The goal is to release a representation useful for a task (e.g., inferring total power consumption of a household) while preventing inference of a private attribute (e.g., household occupancy).
While this should in theory mitigate profiling attacks when the private attribute is the household identity, achieving a good privacy-utility trade-off is challenging~\cite{shateri2019deep}.

\textbf{Matching attacks against smart meter data.} A few matching attacks have been proposed against smart meter data. 
Cleemput et al.~\cite{cleemput2018pseudonymization} showed that aggregate consumption values, e.g., used for billing purposes, can be unique. They are thus able to link the aggregate to its corresponding detailed consumption record available in a large-scale pseudonymized dataset with almost perfect accuracy. 
Tudor et al.~\cite{tudor2013analysis} proposed an attack that can work even when multiple households have the same aggregate consumption values, even for multiple months. 
Jawurek et al.~\cite{jawurek2011smart} argued that anomalous behavior (e.g., information about sick days or working hours) in the household 
can be used to link a real identity to a pseudonymized record, but do not evaluate their hypothesis on a real-world dataset. 
Recently, Voyez et al. \cite{voyez2022unique} studied the uniqueness of smart meter data, showing how consumption records of 3 consecutive days, aggregated at the daily level, are enough to uniquely identify 90\% of 25M individuals. 
Although rounding the data reduces uniqueness, 40\% of individuals are still identifiable when using 7 consecutive days rounded to 3 orders of magnitude. 
Matching attacks, including concurrent work by Voyez et al.~\cite{voyez2022unique}, are complementary to profiling attacks.
Indeed, they allow an attacker to link smart meter data to an identity but are typically prevented by re-pseudonymization strategies used in industry.
Matching attacks that rely on aggregate information, e.g., monthly consumption, require at least data over the same period of time. 
More recent uniqueness-based matching attacks are applicable to data collected over a short period of time but would require such auxiliary data on the person of interest every week. 
Profiling attacks are thus essential to first link back together data about an individual over a long period of time before applying matching attacks.

\textbf{Neural networks for smart meter data.} While neural networks have not, to the best of our knowledge, been used to model smart meter data for user identification, they have been used for electrical load forecasting (ELF)~\cite{breitenbach2022systematic}. 
ELF is a regression task aiming to predict the electrical load consumption of a user at a future timestamp, given their electrical load consumption at previous timestamps. 
The neural network architectures considered for ELF include multilayer perceptrons ~\cite{khotanzad1997annstlf,hayati2007artificial,humeau2013electricity}, recurrent neural networks~\cite{zheng2017electric,bouktif2018optimal,wang2018short,gasparin2022deep}, temporal convolutional networks~\cite{gasparin2022deep,peng2020short}, hybrid networks such as the CNN-LSTM~\cite{alhussein2020hybrid} and the Transformer~\cite{zhao2021short}.
Their success suggests that smart meter data could be, in general, modeled by neural networks. 
The profiling task we study in this paper is however different and the extent to which these methods could generate embeddings of user behavior that are stable across time and can be used for identification is unclear. 

\textbf{Deep learning-based behavioral profiling attacks.} 
A variety of deep learning-based profiling approaches have been developed for different data modalities: identifying an individual from their interactions~\cite{crectu2022interaction} or from an image of their face~\cite{schroff2015facenet}, identifying an author from their writing samples~\cite{fabien2020bertaa}, or a chess player from their actions~\cite{mcilroy2021detecting}. 
The work that is closest to ours is the deep learning-based profiling attack framework for identification across time~\cite{crectu2022interaction}, which profiles a users' weekly interactions for identification across time. 
Differently from this work that used graph attention network embeddings, we focus on neural network architectures for sequential inputs, and study the various dimensions impacting profiling performance in the smart meter domain.

%% file: sections/attack_methodology.tex
\section{Profiling attack design}
\label{sec:attack_methodology}

\subsection{Overview of the approach}
\begin{figure}[t!]
    \centering
    \includegraphics[width=\linewidth]{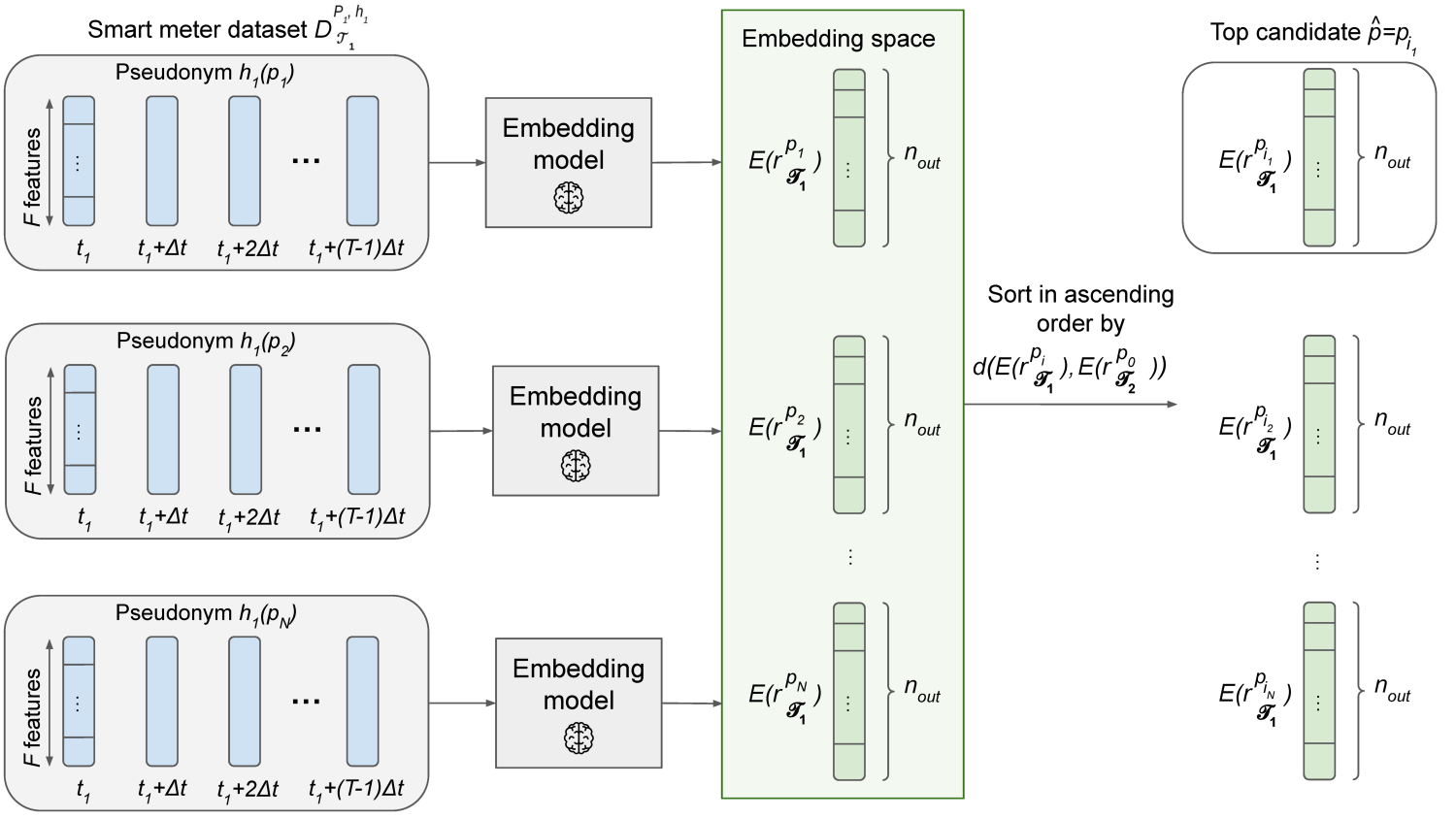}
    \caption{\textbf{Profiling attack pipeline}.
    }
    \label{fig:attack_pipeline}
\end{figure}

In this work, we propose a new deep learning profiling attack against re-pseudonymized smart meter records (Fig.~\ref{fig:attack_pipeline}). 
We train a neural network architecture, referred to as the \textit{embedding model}, to extract features from weekly smart meter records $E(R^p_{\mathcal{T}}) \in \mathbb{R}^d$ with $|\mathcal{T}|=1$ week.
We then use nearest neighbor matching to retrieve the smart meter record from period $\mathcal{T}_1$ that is the most similar to the auxiliary information about a target user $p_0$ coming from period $\mathcal{T}_2$.
The similarity is determined based on the features extracted by the embedding model and the Euclidean distance function $d$:
\begin{equation}
    h_1(\hat{p}) = \arg\min_{h_1(p), p \in P_1} d(E(R^{p}_{\mathcal{T}_1}), E(R^{p_{0}}_{\mathcal{T}_2}))
\end{equation}

Unlike previous hand-engineered feature extraction approaches~\cite{buchmann2013re,tudor2013analysis}, our embedding function is optimized directly for the task of piecing together weekly smart meter records.  
The embedding model is designed to learn what features to extract from a smart meter record that both (1) encode the unique patterns of individuals and (2) are stable across time (see Sec.~\ref{subsec:model_architecture}-~\ref{subsec:model_optimization} for details).

We focus on \textit{weekly} re-pseudonymization strategies, i.e., $\mathcal{T}_1$ and $\mathcal{T}_2$ each span one week.
Indeed, human behavior is naturally organized in weekly patterns.
Weekly data is the minimum frequency allowing smart meter data to retain utility in most use cases, such as analyzing consumption patterns to help users optimize their energy usage and performing marketing campaigns and longitudinal studies of users.
More frequent re-pseudonymization sharply reduces the utility of the data, since the consumption of users varies strongly from one day to the next, e.g., between a weekday and the weekend.
In some cases, re-pseudonymization may need to be even less frequent, e.g., at least two weeks of consumption are needed to perform a natural experiment, one control and one treatment week.
While we here focus on weekly re-pseudonymization, we discuss in Sec.~\ref{sec:discussion} results of our attack against bi-weekly and monthly re-pseudonymization.

We focus on targeted attacks where the goal of the attacker is to re-identify one target user: $P_2 = \{ p_0\} \subset P_1$.
We assume that the attacker has access to an auxiliary dataset $D_{\text{aux}}$ consisting of the smart meter records of an \textit{auxiliary user set} $P_{\text{aux}} \subset \mathcal{P}$ over a time period $\mathcal{T}_{\text{aux}}$ that is disjoint from $\mathcal{T}_1$ and $\mathcal{T}_2$.
While the auxiliary and reference user sets may overlap, the pseudonyms of the common users are assumed to be different. 
We instantiate the profiling attack in two concrete scenarios. 

\underline{Scenario (I)}: In this scenario, the auxiliary user set is the same as the reference user set: $D_{\text{aux}} = D_{\mathcal{T}_{\text{aux}}}^{P_\text{aux}, h_{\text{aux}}}$ with $P_{\text{aux}}=P_1$ and $h_{\text{aux}}=h_1$, i.e., the users have the same pseudonyms in $\mathcal{T}_{\text{aux}}$ and $\mathcal{T}_1$. 
This scenario could arise, e.g., when the attacker gains access to a pseudonymized smart meter dataset as well as to the re-pseudonymized smart meter readings from one of the users ($D_{\mathcal{T}_2}^{P_2, h_2}$). 
We show that an embedding function can be trained on the auxiliary dataset to extract a set of features useful for identifying users across time.

\underline{Scenario (II)}: In this scenario, we assume that the auxiliary and reference user sets are disjoint: $D_{\text{aux}} = D_{\mathcal{T}_{\text{aux}}}^{P_{\text{aux}}, h_\text{aux}}$ with $P_{\text{aux}} \cap P_1 = \emptyset$.
Thus, the attacker aims to identify the correct match of the target user $p_0$ in $D_{\mathcal{T}_1}^{P_1, h_1}$  
using one week of auxiliary information $R_{\mathcal{T}_2}^{p_0}$ but \textit{no other information}.
From a threat modeling perspective, this scenario is more realistic as only one weekly sample is available from the reference user set (including the target user), in $\mathcal{T}_1$.
We refer to this scenario as \textit{one-shot identification}.
In this scenario, we show that the features learned by an embedding model trained on users in $D_\text{aux}$  can be ``transferred'' from the auxiliary user set to the disjoint reference user set at almost no cost in accuracy.

\subsection{Neural network embedding model}
\label{subsec:model_architecture}
As our embedding model, we adapt six neural networks that have been proposed for smart meter load forecasting to the task of identification. 
These cover a wide variety of architectures that can operate on time-series inputs: a Multilayer Perceptron (MLP), two Recurrent Neural Network (RNN) architectures; a Long-Short Term Memory Network (LSTM) and a Gated Recurrent Unit (GRU), as well as a Time Convolutional Network (TCN), an CNN-LSTM, and a Transformer. 
Note that our goal is not to invent a novel architecture for smart meter modeling, but rather to adapt architectures that have been proposed for load forecasting to the identification task and to evaluate the robustness of re-pseudonymization strategies. 
In particular, we evaluate the robustness of these strategies against moderately capable attackers relying on state-of-the-art architectures for sequential modeling that are readily available in (or easy to extend from) open-source libraries.
The different architectures we consider also help us understand the risk posed by attackers with different capabilities, e.g., an MLP is relatively quick to train even on a CPU while a Transformer is much larger and may require extensive hyperparameter tuning.

We describe how each architecture processes a weekly smart meter record.
All the architectures take as input the records of a user in a week  $\mathcal{T}=[t_1, t_1')$, with $t_1'=t_1 + T \times \Delta t$:
\begin{equation}R_{\mathcal{T}} = ( r_{t_1} \;,\; r_{t_1+\Delta t} \;,\; \ldots \;,\; r_{t_1+ (T-1) \times \Delta t} )
\end{equation}
where we have dropped the user superscript.
As a reminder, the total number of features available is $T \times F$. 
As useful notation for architectures operating sequentially on daily features, we denote by $t$ the duration of a day in the same unit as the time granularity $\Delta t$, so that the number of features in a day is equal to $\lfloor \frac{t}{\Delta t}\rfloor$. We will also denote by $t_d$ the first timestamp in day $d$, so that the values of feature $f$ recorded in day $d$ are equal to:
\begin{equation}R_{d,f} = \big(r_{t_d, f} \;,\; r_{t_d + \Delta t, f} \;,\; \ldots \;,\;r_{t_d + (\lfloor \frac{t}{\Delta t}\rfloor-1) \Delta t, f} \big),\ f=1,\ldots,F\label{eq:daily_features}
\end{equation}
Finally, we denote by $n_{\text{out}}$ the size of the output embedding.

\begin{figure}[t!]
    \centering
    \includegraphics[width=0.7\linewidth]{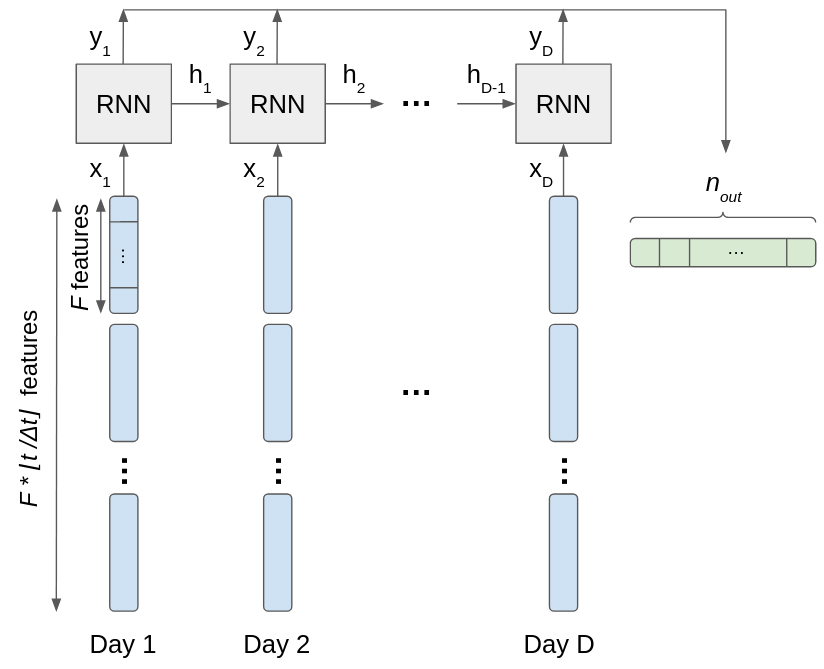}
    \caption{\textbf{Using RNNs for feature extraction}.
    }
    \label{fig:rnn_input}
\end{figure}

\textbf{Multilayer Perceptron (MLP).} 
The MLP is a standard neural network consisting of one or more linear transformations, each followed by a non-linear activation function.
In spite of its simplicity, the MLP can approximate a broad range of functions, a result known as the universal approximation theorem~\cite{hornik1989multilayer}.
We apply the MLP to the flattened list of features values, in temporal order:
\begin{equation}
(r_{t_1, 1} \;,\; \ldots \;,\; r_{t_1, F} \;,\; \ldots \;,\; r_{t_1 + (T-1)\Delta t , 1} \;,\; \ldots \;,\; r_{t_1 + (T-1)\Delta t, F})
\end{equation}

\textbf{Recurrent Neural Network (RNN).} RNNs are a well-established family of architectures designed to operate on sequential inputs. 
In the electrical load forecasting literature (see Sec.~\ref{sec:related_work} for an overview), inputs are modeled as fine-grained sequences, where each element consists of the $F$ features recorded at a particular timestamp. 
To take into account the circadian rhythm of human behavior, we use sequences of daily features from $D$ days: $x_1, \ldots, x_D$, where $x_d=[R_{d,f}]_{f=1,\ldots,F}$  ($[\cdot]$ denotes concatenation).
The RNN consists of $L$ layers.
A layer $l$ embeds the elements of the input sequence $x_1^l, \ldots, x_D^l$ (setting $x_d^1=x_d$), in order, and outputs a hidden state at each step $d$, $h^l_d=E(x^l_1, \ldots, x^l_d, h^l_{d-1})$.
We implement the two most widely used RNN architectures, Long Short-Term Memory (LSTM)~\cite{hochreiter1997long} (where the hidden state is decomposed into two vectors, an output state $y^l_d$ and a cell state $c^l_d$: $h^l_d=(y^l_d, c^l_d)$ having different functions in the computation) and the Gated Recurrent Unit (GRU)~\cite{cho2014properties} (where $h^l_d = y^l_d$ to simplify the computation).
The output sequence of the previous layer is used as the input sequence to the next layer: $x_d^l = y_d^{l-1}, l=2, \ldots, L$.
We use the sum of the output states after each step of the last layer $L$ as our embedding: $E(R_{\mathcal{T}})=y^L_1 + \ldots + y^L_D$.
Fig. \ref{fig:rnn_input} illustrates the architecture for $L=1$. 

\begin{figure}
    \centering
    \includegraphics[width=\linewidth]{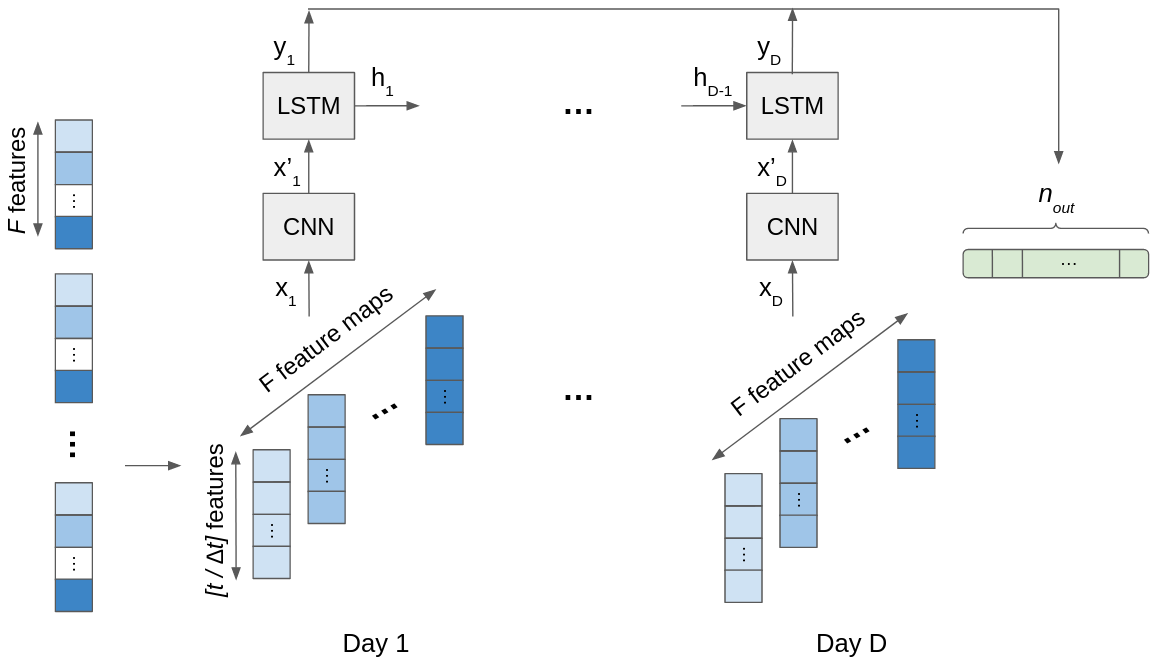}
     \caption{\textbf{Using a CNN-LSTM for feature extraction}.
     }
    \label{fig:cnnlstm_input}
\end{figure}

\textbf{Convolutional Neural Network-Long Short-Term Memory Network (CNN-LSTM).} 
CNN-LSTM have obtained state-of-the-art results in load forecasting~\cite{alhussein2020hybrid}. 
The model we use consists of $L$ layers, each composed of two parts: a CNN part for feature extraction and an LSTM part for sequence learning. 
Like the RNN, the CNN-LSTM  processes the daily features sequentially. 
However, the features in a day are  grouped by their type into $F$ feature maps before they are processed by the CNN.
The feature maps are given by Eq.~\ref{eq:daily_features} and their
 size is equal to $\lfloor\frac{t}{\Delta t}\rfloor$. 
We design the CNN part using a combination of convolutional layers, for inter-feature modelling, and max pooling layers, for feature downsampling: $(x'_d)^l=\text{CNN}(x_d^l)$ (we refer the reader to Sec.~\ref{subsec:architecture_details} for the complete details). 
The LSTM part operates on the sequence of daily features extracted by the CNN, as described previously. 
Similarly, we use the sum of the output states after each step of the last layer $L$ as our embedding: $y_1^L + \ldots + y_D^L$ as our embedding. 
Fig. \ref{fig:cnnlstm_input} illustrates the architecture for $L=1$. 

\textbf{Temporal Convolutional Network (TCN).} The TCN architecture was found to achieve higher performance for long-term sequential modelling compared to LSTMs or GRUs~\cite{bai2018empirical} and has been explored in the load forecasting literature~\cite{peng2020short}. 
The TCN encodes information from long sequences by aggregating features in a tree-like fashion. 
Specifically, it applies one or more layers of temporal blocks of dilated convolutions to the temporal sequence. 
The depth of the TCN, i.e., its number of layers and the kernel size, control the complexity of the network. 
We use the original implementation made available by its authors~\cite{bai2018empirical} and model our input data as a temporal sequence of daily features. 
The features available on each day are organized into a number of $\lfloor t / \Delta t \rfloor F$ channels, each corresponding to the consumption at a specific timestamp (e.g., 11AM).
In our experiments, we found one layer and a kernel size equal to the number of days in the sequence $D$ to perform better than using more layers and smaller kernel sizes. 
We believe that this is due to our sequences being short (of $D=7$ days) and the time granularity being high ($\Delta t=1$ hour). 
For each day in the sequence $1 \leq d \leq D$, a convolutional layer is applied to features from days 1 to $d$, yielding an embedding of size
$n_{\text{out}}$. 
We use the sum of the $D$ outputs as our embedding. 
We refer the reader to Fig.~\ref{fig:tcn_input} in the Appendix for an illustration of our adapted TCN architecture. 

\textbf{Transformer.} The Transformer architecture was introduced as a simpler alternative to RNNs for sequence modeling tasks~\cite{vaswani2017attention}. It forms the basis of state-of-the-art language models used in natural language processing tasks such as keyboard auto-completion.
We here implement a Transformer operating on the sequence of daily features, like the other architectures.
It consists of $L$ encoder layers, each composed of a self-attention mechanism and a feedforward layer.
The self-attention takes as input the $D$ output encodings of the previous layer (the daily features for $L=1$) and produces $D$ encodings.
The $d$\textsuperscript{th} encoding is obtained by linearly combining (a linear projection of) the inputs with weights given by their relevance to the $d$\textsuperscript{th} input.
More specifically, the inputs $x_1, \ldots, x_D$ are projected to query, key, and value vectors $x^q_d =x_d Q$,  $x^k_d = x_d K$, and $x^v_d = x_d V$, and the output encoding is computed as $\sum_{d'=1}^D a(d, d') \times x^v_{d'}$, where $a(d, d') \propto (x^q_{d})^T (x^k_{d'})$.
We refer the reader to Vaswani et al.~\cite{vaswani2017attention} or to the PyTorch library implementation for complete details.
Then, a feedforward layer is applied to each output encoding before passing it as input to the next layer.
We use the sum of output encodings at the last layer as our embedding.

\subsection{Optimizing the embedding model}\label{subsec:model_optimization}

We use the triplet loss for optimizing the parameters of the embedding model in order to extract features that can be used for identification across time. 
The triplet loss was introduced by Schroff et al.~\cite{schroff2015facenet} for the purpose of facial recognition and has yielded state-of-art results in profiling attacks against interaction data~\cite{crectu2022interaction}. 
Differently from load forecasting which is framed as a regression task with loss objectives such as the mean absolute error~\cite{alhussein2020hybrid} and previous profiling attacks framed as a supervised learning task ($N$-way classification)~\cite{jawurek2011smart}, we frame large-scale identification as a self-supervised learning task, via the triplet loss objective function.

The goal of the triplet loss is to bring the embeddings of the same user closer to each other in the embedding space while pushing them away from embeddings of other users.
We define a sample as the smart meter record of a user over one week, similarly to Cre\c{t}u et al.~\cite{crectu2022interaction}.
The triplet loss is computed over three samples, two from the same user $p$ and different time periods $\mathcal{T}_A \neq \mathcal{T}_B$ (the \textit{anchor} and \textit{positive samples}) and a third sample from another user $p' \neq p$ and any time period $\mathcal{T}_C$ (the \textit{negative sample}): $\mathcal{L}(p, \mathcal{T}_A, \mathcal{T}_B, p', \mathcal{T}_C) = \max\Big(0,  d\big(E(R^{p}_{\mathcal{T}_A}), E(R^{p}_{\mathcal{T}_B}) \big)  - d\big(E(R^{p}_{\mathcal{T}_A}), E(R^{p'}_{\mathcal{T}_C})\big) + \lambda\Big)$,
where $\lambda$ is a margin parameter. 
We refer the reader to Appendix~\ref{appendix:triplet_loss} for a detailed explanation of the triplet loss.

We optimize the parameters of the neural network using mini-batch gradient descent. 
We generate triplets for the profiling task similarly to Cre\c{t}u et al.~\cite{mcilroy2021detecting}.
More specifically, we define an epoch as a complete pass through all the users $P$ in the dataset  $D_{\text{aux}}$. 
In each batch of users (of fixed size $B$), we sample a day $d$ uniformly at random among the days available in the time period $\mathcal{T}_{\text{aux}}$. 
We use the smart meter record of users in the batch and the week starting at $d$ as the anchor samples.
For each anchor sample, we use two positive samples and a negative sample, thus generating two triplets.
The positive samples comes from the weeks starting at days $d-l$ and $d+l$, respectively, where $l$ denotes a lag parameter. 
The day addition and subtraction operations are computed modulo the number of days in the dataset. 
A lag of $l=7$ thus corresponds to sampling from disjoint weeks, while a lower lag value corresponds to sampling from overlapping weeks. 
As noted in previous work on the triplet loss~\cite{schroff2015facenet,hermans2017defense}, the selection of negative samples plays a crucial role in efficiently optimizing the embeddings. 
Here we use online hard mining~\cite{schroff2015facenet}, where the negative sample is the closest embedding in the batch that does not belong to the same user.

%% file: sections/experimental_setup.tex
\section{Experimental setup}\label{sec:experimental_setup}
In this section, we present our evaluation setup for profiling attacks against a real-world smart meter dataset. 

\subsection{Dataset}
We evaluate our attack against a dataset of smart meter records provided by EDF UK in a secure environment for the purpose of this study.
A contract, reviewed by our Data Protection Officer, was signed between EDF UK and Imperial College London for the sharing of the dataset in pseudonymized form for the purpose of this study, along with security measures requirement. The dataset does not contain personally identifiable information and no individual was re-identified as part of this research. 
The results of this paper have been shared with the data provider, EDF UK, who approved for the results to be made available in this academic publication. The dataset used in the study cannot be made available for privacy and contractual reasons.

The dataset consists of pseudonymized electricity and gas consumption records from individual properties collected at a time granularity of $\Delta t=1$ hour. 
The dataset comprises two parts. 

The first part, which we use in most experiments, consists of the electricity and gas consumption records of $N=5139$ users $P$ over 49 consecutive weeks. 
We assume that an attacker has access to an auxiliary dataset $D_{\text{aux}}$ consisting of the records of all the users (i.e., $P_{\text{aux}}=P$) from the first $M=40$ weeks, $\mathcal{T}_{\text{aux}}$.
We use the auxiliary dataset to train and tune the hyperparameters of a deep learning-based profiling model $E$.
In scenario (I), we aim to identify target users from $P_{\text{aux}}$, i.e., $P_1 = P_{\text{aux}}$ between two disjoint weeks: the 41-st week ($\mathcal{T}_1$) and a subsequent week ($\mathcal{T}_2$).
More specifically, to vary the time gap between $\mathcal{T}_1$ and $\mathcal{T}_2$, we use weeks $42$ to $49$ in turn as target week $\mathcal{T}_2$ from which auxiliary information about the target user is drawn. 

The second part of the dataset consists of the electricity consumption records of 62170 additional users $P'$ from reference week 41 ($\mathcal{T}_1$) and the target week 42 ($\mathcal{T}_2$).
We use this dataset to first evaluate our model on scenario (II), i.e., on a disjoint set of users where $P_1 \cap P_{\text{aux}} = \emptyset$, by randomly selecting a subset $P_1 \subset P'$ of size $N=5139$ (Sec.~\ref{subsec:model_applicability_unseen}).
Second, we use it to test the scalability of the attack on large population sizes, by evaluating it on a reference set $P_1=P' \cup P_{\text{aux}}$ (i.e., a total of 67309 reference users, Sec.~\ref{subsec:increasing_population}).

We pre-process the dataset by (1) replacing multiple recordings reported for the same timestamp with the average value, (2) filling the consumption values for missing timestamps with zeros, and (3) clipping outlier values in the gas dataset 
We refer the reader to Appendix~\ref{appendix:data_preprocessing} for details.

\subsection{Attack success metric}\label{subsec:attack_metric}
Our goal is to evaluate how well our profiling attack is able to re-identify users based on their re-pseudonymized smart meter records in weeks $\mathcal{T}_1$ and $\mathcal{T}_2$.
Towards this goal, we perform our targeted attack against every user $p_0\in P_1$, computing the distance between the embedding of the target user in week $\mathcal{T}_2$, $E(R^{p_0}_{\mathcal{T}_2})$, and the embeddings of all the reference users $p_1, \ldots, p_N \in P_1$ in week $\mathcal{T}_1$: $E(R^{p_1}_{\mathcal{T}_1}), \ldots, E(R^{p_N}_{\mathcal{T}_1})$.
We re-order the reference users (i.e., their pseudonyms) decreasingly by their distance to the target user: 
\begin{equation}
d\big( E(r^{p_{i_1}}_{\mathcal{T}_1}), E(r^{p_0}_{\mathcal{T}_2})\big) \leq   
\ldots 
\leq d\big( E(r^{p_{i_N}}_{\mathcal{T}_1}), E(r^{p_0}_{\mathcal{T}_2})\big)
\end{equation}
Our metric for attack success is  the \textit{probability of identification within rank $R$}, defined as the percentage of target users $p_0$ in $P_1$ retrieved among the top $R$ users, i.e., $\exists k \in \{ 1, \ldots, R\}$ such that $p_0 = p_{i_k}$.
This metric is standard in the re-identification literature.

\subsection{Model architecture and training details}\label{subsec:architecture_details}
We present the architecture details for the neural networks implemented in this paper and details about the training process.

\textbf{General parameters.}  We use  embeddings of size of $n_{\text{out}}=32$.
We scale them by their Euclidean norm to have a norm of 1, with the exception of the Transformer whose performance is negatively impacted considerably by this choice.
We leave for future work the study of potential causes of this behavior in Transformers.
As the smart meter dataset records consumption on an hourly basis ($\Delta t$=1 hour), there are 168$\times F$ features in a weekly smart meter record. In the experiments, $F=1$ when using electricity consumption records and $F=2$ when using both electricity and gas. 
We scale the features to the $[0, 1]$ range before passing them as input to the model.

\textbf{Architecture details.} The MLP architecture is composed of three hidden layers of sizes 128, 64 and 32, respectively and the non-linear activation used is the Rectified Linear Unit (ReLU). The RNN architectures (LSTM and GRU) both use a hidden size equal to $n_\text{out}$ and we perform hyperparameter search for the number of layers $L \in \{2,3\}$. The CNN-LSTM architecture applies two convolutional layers with number of output channels equal to 16 and 32, respectively and kernel sizes equal to 3 and 4, respectively. The convolutional layers are followed by batch normalization, LeakyReLU activation with a negative slope of 0.01, and a max pooling layer with a kernel size of 2. We perform hyperparameter search for the number of layers $L \in \{1, 2\}$. We use a Transformer architecture that maps the input features to a space of size 128, then applies one or more multi-head attention layers~\cite{vaswani2017attention} with 4 attention heads and an embedding dimension of 64, and finally applies a linear transformation to the output space of size $n_{\text{out}}$. We perform hyperparameter search for the number of layers $L \in \{2, 3\}$. We implemented all the architectures in PyTorch 1.4.0 and all the other choices are the default provided by the library.

\textbf{Training setup.} We train each model on the auxiliary dataset. 
First, we tune (if applicable) the model and training hyperparameters based on the probability of identification within rank $R=1$ for users in $P_{\text{aux}}$ computed on weeks $M-1$ and $M$ of the model trained on the first $M-4$ weeks and early stopped based on the probability of identification within rank $R=1$ for the same users computed on validation weeks $M-3$ and $M-2$. 
Second, a model using the best hyperparameters is trained from scratch on the first $M-2$ weeks and early stopped based the probability of identification within rank $R=1$ computed on validation weeks $M-1$ and $M$. 
We report the probability of identification within rank $R$ obtained using this model on test weeks $M+1$ ($\mathcal{T}_1$) and $M+1+G$ ($\mathcal{T}_2$), for varying gaps $G=1, \ldots, 8$, averaged over 10 runs. 
We use a time gap of one week $G=1$ in all the experiments except for one, where we study the impact of the time gap $G$ (Sec.~\ref{sec:discussion}). 

We refer the reader to Appendix~\ref{appendix:training_details} for details of the training pipeline and hyperparameter search.

\subsection{Baselines}\label{subsec:baselines}
We compare the performance of our proposed approach with six baselines. 
The first two baselines are simple approaches.
First, the \textbf{random guess} assigns an identity uniformly at random among the $N$ identities. 
The probability of identification within rank $R$ is thus equal to $\frac{R}{N}$, for $R = 1, \ldots, N$.
Second, the \textbf{$\mathbb{L}_2$-Raw features} baseline performs, like our approach, nearest neighbor matching to retrieve the smart meter record from $\mathcal{T}_1$ that is the most similar to the target user's record in $\mathcal{T}_2$, as per the Euclidean distance. 
However, differently from our approach, this baseline uses the identity embedding $E(R_{\mathcal{T}}^p)=R_{\mathcal{T}}^p$ after standardizing each feature using the mean and standard deviation estimated on the training dataset.
It thus acts as a strong baseline informative of the \textit{true advantage} of using embedding models for feature extraction.

The remaining four baselines, namely \textbf{Buchmann}~\cite{buchmann2013re}, \textbf{Tudor}~\cite{tudor2015study}, \textbf{Jawurek}~\cite{jawurek2011smart}, and \textbf{Faisal}~\cite{faisal2015quantity} are methods from the literature on profiling attacks against smart meter data. 
In  Appendix~\ref{appendix:literature_baselines}, we describe in detail these baselines and how we implemented them in order to ensure a fair and systematic comparison with our approach.

%% file: sections/results.tex
\section{Results}
\label{sec:results}

\subsection{Scenario (I): model comparison}
\label{subsec:model_comparison}

We first compare the performances of models on scenario (I), in which the reference user set is the same as the auxiliary user set whose data is used to train the model: $P_1=P_{\text{aux}}$. 
We compute the probability of identification within rank $R=1 \;,\; \ldots \;,\; N=5139$ on the test weeks $M+1$ ($\mathcal{T}_1$) and $M+2$ ($\mathcal{T}_2$) that are disjoint from the time period $\mathcal{T}_{\text{aux}}$ used to train the model.
For every architecture, we train 10 models and report the mean with standard deviation (std) of their performance.
For a fair comparison with the existing literature, we only use electricity records in this experiment.

\begin{figure}[h]
    \centering
    \includegraphics[width=\linewidth]{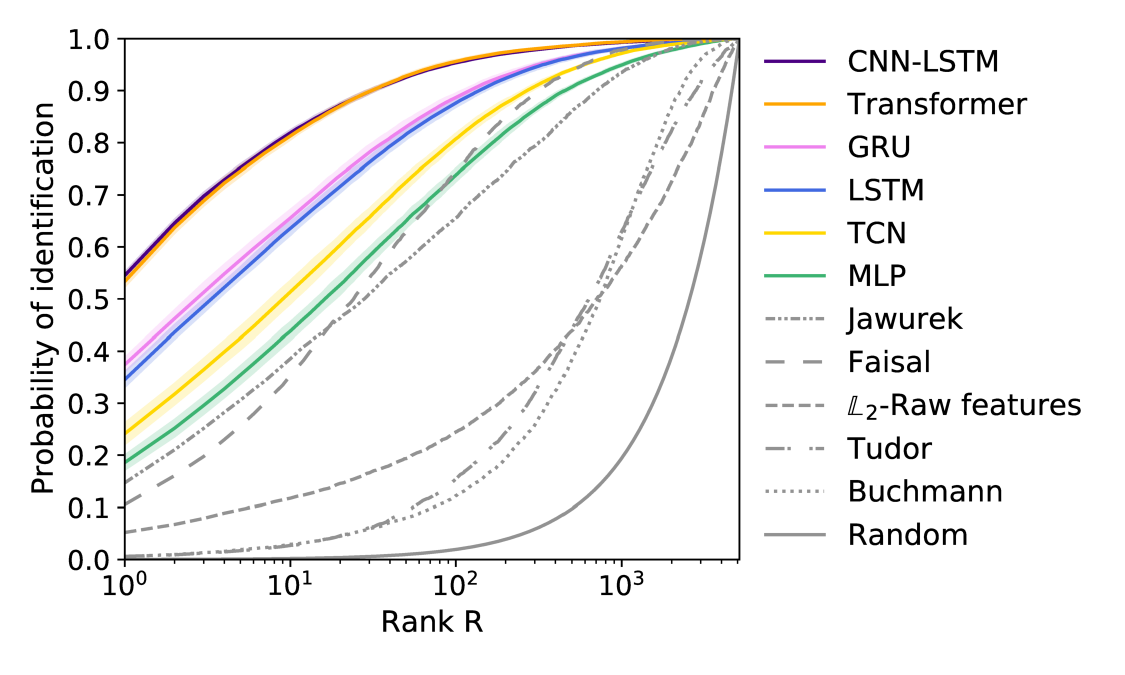}
    \caption{Scenario (I): Attack performance using different approaches. 
    For the embedding-based approaches, we report the mean with standard deviation over 10 runs.
    All our embedding approaches are superior to the baselines, with the CNN-LSTM and the Transformer vastly outperforming them.
    }
    \label{fig:model_performances_gap1}
\end{figure}

Figure \ref{fig:model_performances_gap1} shows that our adaptation of the CNN-LSTM architecture identifies users from a week of electricity consumption records $54.5\%$ of the time, strongly outperforming baselines on their task (scenario (I)). 
It also shows that the probability of identification increases for higher ranks, as users can be retrieved $75.2\%$ of the time among the top 5 candidates and $81.9\%$ of the time among the top 10 candidates. 
Interestingly, despite using more than four times as many parameters (100k vs. 23k), the Transformer architecture is on par with the CNN-LSTM. 
More specifically, it achieves a probability of identification within rank 1 of $53.4\%$, and of $74.6\%$ and $81.2\%$, within ranks 5 and 10, respectively. 

The next best architecture, the GRU, is performs significantly worse, correctly identifying users uniquely $37.4\%$ of the time. This performance corresponds to a loss of 17.1 p.p. compared to the CNN-LSTM. As the LSTM part of the CNN-LSTM architecture is similar to the GRU, we hypothesize that this drop is due to the absence of the CNN part which transforms the daily input features before they are passed through the recurrent part. 
To test this hypothesis, we modified the GRU architecture so that the daily input features are mapped linearly to a space of same size as the output of the CNN part (namely 128). While this increases the number of parameters of the GRU architecture from 12k to 25k, 2k larger than that of the CNN-LSTM architecture, we only observe a small performance gain of 0.3\%. The other RNN architecture, the LSTM, achieves a slightly lower performance compared to the GRU and it is able to correctly identify users $34.6\%$ of the time. 

Next, the TCN achieves a performance of $24.1\%$ and it is closely followed by the MLP, at $18.6\%$. 
It is interesting to note that the TCN architecture performs 12.5\% lower than the RNN architectures. The TCN architecture is indeed generally highly regarded in the literature for its capacity to process time series \cite{bai2018empirical}. We believe that the TCN could, in principle, perform better if applied to longer sequence lengths (of more than 7 days), but this falls outside of our attack model and of the scope of this work.

Finally, our adapted CNN-LSTM and actually all our deep learning models strongly outperforms the previous methods on the standard task in the literature. The best performing baseline, by Jawurek et al.~\cite{jawurek2011smart}, is able to correctly identify users $14.7\%$ of the time, 39.8 p.p. lower than our CNN-LSTM. Furthermore, this baseline can be inefficient as it requires to train $N=5139$ classifiers to distinguish between each user and the rest. 
In contrast, our embedding-based approach is able to directly learn a user's profile that distinguishes them from the others. The second best performing baseline is the one proposed by Faisal et al.~\cite{faisal2015quantity} and only achieves a probability of identification within rank 1 of $10.5\%$. Last, the baselines by Buchmann et al.~\cite{buchmann2013re} and Tudor et al.~\cite{tudor2015study} achieve the lowest performance. They both correctly identify users $0.6\%$ of the time. We note that this is one order of magnitude better than the performance of a random guess approach, $0.02\%$. While the approach of Tudor et al. significantly simplified the one of Buchmann et al~\cite{buchmann2013re} and outperformed it on a dataset of 400 users~\cite{tudor2015study}, it achieves the same performance when applied to our dataset of $N=5139$ users.  These results show that the existing literature strongly under estimates the general risk of re-identification of smart meter data.

Interestingly, the simple $\mathbb{L}_2$-Raw features approach is surprisingly robust: it correctly identifies users $p_1=5.2\%$ of the time and $p_{10}=11.7\%$ of the time among the top 10 candidates. The gap between our method and this baseline demonstrates the advantage of using embeddings to automatically extract features tailored for the profiling task. 
Furthermore, this suggests that a small, non-negligible fraction of the users in the dataset have a highly regular consumption behavior from one week to the next that is also very unique. Finally, it puts into perspective results of prior works. 

\begin{table}[t!]
\centering
    \caption{Comparison between attack performances (mean with std over 10 runs) in scenarios (I) and (II).
    }
    \label{tab:unseen_userset}
    \begin{tabular}{cccc}
    \toprule
     \textbf{Model} &
     \textbf{(I) $P_1=P_{\text{aux}}$} & 
     \textbf{(II) $P_1 \cap P_{\text{aux}} = \emptyset$} & \textbf{Diff.} \\ 
     \midrule
     CNN-LSTM & 54.5 (0.6) & 52.2 (0.6) & 2.3 \\
     Transformer & 53.4 (1.0) & 51.3 (0.8) & 2.1 \\ 
     GRU & 37.4 (1.3) & 36.2 (1.5) & 1.2 \\ 
     LSTM & 34.6 (1.4) & 33.3 (1.0) & 1.3 \\  
     TCN & 24.1 (2.0) & 23.8 (2.1) & 0.4 \\ 
     MLP & 18.6 (1.4) & 18.8 (1.3) & -0.2\\ 
     $\mathbb{L}_2-$Raw features & 5.2 (-) & 6.3 (-) & -1.1 \\ 
     \bottomrule
    \end{tabular}
\end{table}

\subsection{Scenario (II): applicability to unseen users}
\label{subsec:model_applicability_unseen}
Scenario (I) reflects the traditional task in the literature: all the machine learning-based methods so far require the attacker to have a large amount of data about the user of interest to be able to re-identify them. 
This data is not available to the attacker when re-pseudonymization strategies are used before releasing the data, leaving us with only ancillary evidence to evaluate the effectiveness and overall trade-off achieved by re-pseudonymization. 

In contrast, our method can be used to re-identify users using only one week of data. 
More specifically, we show how our approach can identify users unseen during training.
The embedding function is transferrable to other users because it learns to map a weekly record to a set of features tailored to identification across time.
This makes the attack applicable to frequent re-pseudonymization risk mitigation strategies. 

We test the applicability of our model to scenario (II), corresponding to \textit{one-shot identification}, in which the goal is to identify a target user among reference users \textit{unseen during training}. 
The model is trained on the same dataset from user set $P_{\text{aux}}$ as before.
However, in this scenario, the reference user set $P_1$ is disjoint from $P_{\text{aux}}$ and only one sample is available from the reference users, from week $M+1$ ($\mathcal{T}_1$).
Then, given a target user $p_0$, we assume as before the attacker to have access to their smart meter record from week $M+2$ ($\mathcal{T}_2$), and the goal is to link this record with the correct identity in the reference user set from week $\mathcal{T}_1$.
In our evaluation, we sample a disjoint user set $P_1 \cap P_{\text{aux}} = \emptyset$, of equal size $|P_1|=|P_{\text{aux}}|=5139$. 

Table \ref{tab:unseen_userset} shows that all our models can accurately identify unseen user sets across weekly re-preseudonymized datasets, losing at most 2.3 p.p. in performance. 
The best performing architectures achieve a probability of identification within rank 1 of $52.2\%$ (CNN-LSTM) and $51.3\%$, respectively. We observe a slight drop in performance compared to the previous scenario, of 2.3 p.p. for the CNN-LSTM and of 2.1 p.p. for the Transformer architecture. 
The small difference in performance could be due to the high expressiveness of these architectures, allowing them to memorize details of the training users that do not generalize to other users. The models that perform less well, namely the GRU, the LSTM, the TCN, and the MLP do not yield a significant drop in performance. This suggest that the profiles extracted by these models are quite robust, even though they are less expressive of the differences between the users. 

\subsection{Increasing the population size}
\label{subsec:increasing_population}

The second strong limitation the existing literature suffers from is the small datasets on which results were reported on so far. Indeed, with re-identification accuracy expected to decrease with dataset size, it was unclear the extent to which the fairly high accuracy numbers reported in previous work were applicable to the large to very large datasets currently used in practice.

We study the applicability of our models to match users in re-pseudonymized weeks $M+1$ ($\mathcal{T}_1$) and $M+2$ ($\mathcal{T}_2$), varying the size of the reference user set $P_1$ (which we refer to as the population size) from 2 to 67309.
For the large majority of the users we evaluate on, only one week of data is made available to the model, from the reference time period $\mathcal{T}_1$, at test time. 
Specifically, $62170=67309-5139$ of the users are unseen by the model during training.
We compute the probability of identification within rank 1 of our model when applied to population sizes of 2, 100-1k using a step size of 100, 1k-10k using a step size of 1k, 10k-65k using a step size of 5k and finally 67309. 

\begin{figure}[t!]
    \centering
    \includegraphics[width=0.9\linewidth]{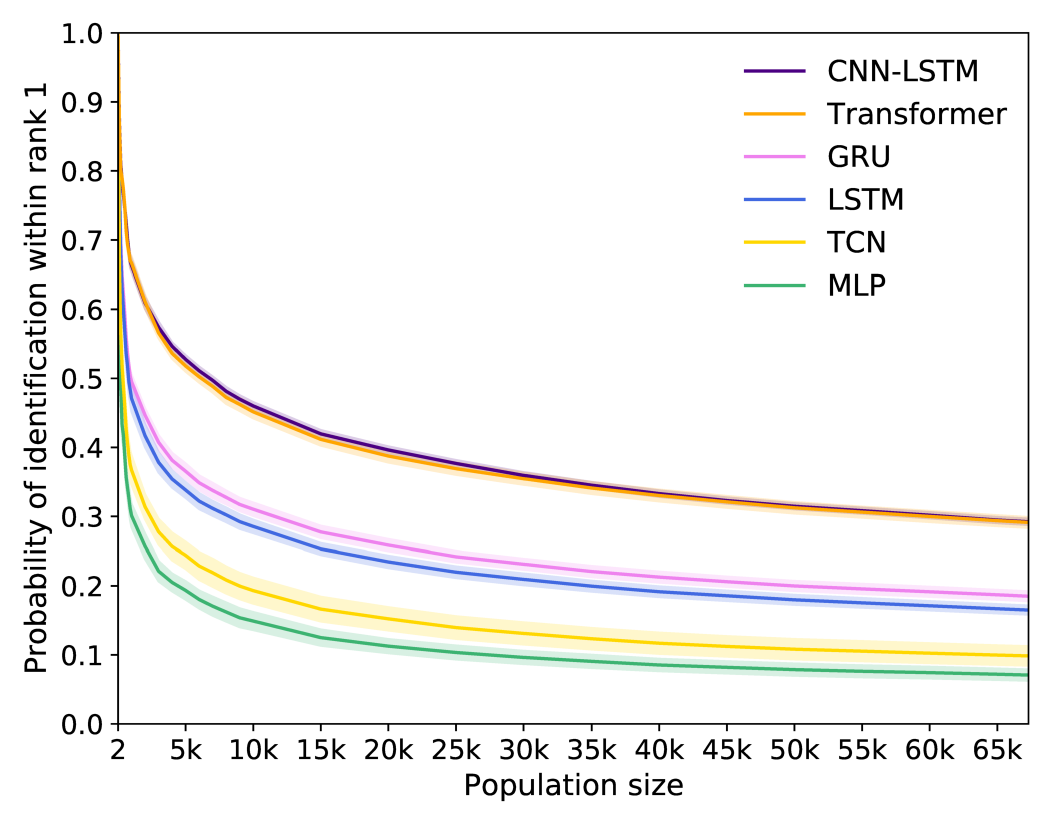}
    \caption{Attack performance (mean with std over 10 runs) when increasing the population size. 
    }
    \label{fig:incr_pop_size}
\end{figure}

Fig.~\ref{fig:incr_pop_size} shows that, as expected, the probability of identification within rank 1 computed using our models decreases as the number of users varies from 2 to 67k. While sharp at first, the decrease slows down significantly after 10k. For instance, the CNN-LSTM accuracy decreases by 20.3\% from $|P_1|=1$k to $|P_1|=10$k but only by 6.4\% from $|P_1|=10$k to $|P_1|=20$k. The CNN-LSTM is indeed still able to identify $29.2\%$ of users in a dataset of 67309 users. 

\textbf{Summary.} Taken together, our results show that frequent re-pseudonymization strategies are ineffective even in very large datasets. 
Our approach strongly outperforms previous work at the traditional task considered in the literature (scenario (I)). 
While the risk from previous work was mitigated by frequent re-pseudonymi- zation strategies, our one-shot identification approach is not. 
Finally, we show for the first time how profiling attack are a risk even against the large-scale datasets currently collected and shared.

\begin{figure}[t!]
\centering
    \includegraphics[width=\linewidth]{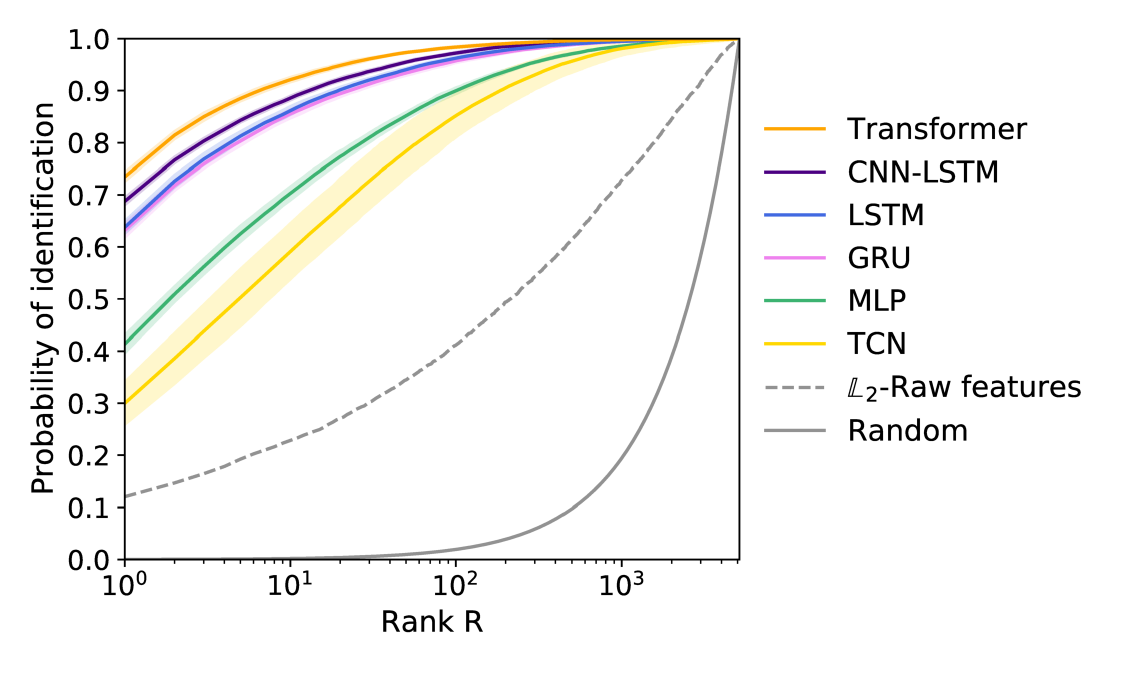}
\caption{\textbf{Attack performance (mean with std over 10 runs) when combining electricity and gas consumption.} 
}
\label{fig:elgas}
\end{figure}

\subsection{Combining electricity and gas consumption}
\label{subsec:features_influence_attack}
We have so far considered the scenario when the attacker has access to (and aims to piece together) electricity consumption records. 
We made this choice as we wanted to perform a fair comparison with previous works which have focused exclusively on electricity. 
We now consider the scenario where the attacker has access to both electricity and gas records. 
Smart meters indeed often record both the electricity and gas consumption of households~\cite{ukreportsmartmeterstats}. 
We retrain the models on both electricity and gas consumption features using the best hyperparameters from Sec.~\ref{subsec:model_comparison}. 
We evaluate the performance of the models on the same test weeks as before. 

Fig.~\ref{fig:elgas} shows that jointly modelling the electricity and gas consumption increases the attack performance considerably. Very interestingly, now that more data is available, the Transformer architecture significantly outperforms the CNN-LSTM. It indeed successfully identifies $73.4\%$ of 5,139 users while the CNN-LSTM only identifies $68.8\%$ of them. These results show that when more data is available the extra parameters of the Transformer allows it to outperform other approaches.
The other methods achieve $63.8\%$ for the LSTM, $63.2\%$, for the GRU, $41.4\%$ for the MLP, $30.0\%$ for the TCN and finally $12.0\%$ for the $\mathbb{L}_2$-matching.

%% file: sections/discussion.tex
\section{Discussion}
\label{sec:discussion}

We now discuss in more detail several assumptions we have made. 

\textbf{Frequency of re-pseudonymization.}
Throughout this paper, we have focused on weekly re-pseudonymization, assuming that $\mathcal{T}_1$ and $\mathcal{T}_2$ both span one week, i.e., $|\mathcal{T}_1|=|\mathcal{T}_2|=1$ where we use $|\cdot|$ to denote the length of the period in weeks.
In some use cases, it may be necessary to re-pseudonymize data less frequently, e.g., on a biweekly basis to perform a natural experiment using one control and one treatment week.
We evaluate the ability of our approach to link together smart meter records re-pseudonymized after $F\in \{1, 2, 4 \}$ weeks, setting $|\mathcal{T}_1|=|\mathcal{T}_2|=F$ weeks with $\mathcal{T}_1 \cap \mathcal{T}_2=\emptyset$.
We do this by computing for every reference user the average embedding of their weekly records in $\mathcal{T}_1$ and for the target users the average embedding of their weekly records in $\mathcal{T}_2$.
Then, as before, we retrieve for every target user their nearest neighbor in the reference user set using the Euclidean distance between the embeddings.
We design this experiment using all the 9 weeks of test data (i.e., that are not used to train the models) by generating all the pairs of consecutive periods $\mathcal{T}_1$ and $\mathcal{T}_2$ of $F$ weeks.
For $F=1$, there are 8 such pairs (week 1 vs. 2, 2 vs. 3, up to 8 vs. 9) while for $F=4$ there are 2 such pairs (weeks 1-4 vs. 5-8 and 2-5 vs. 6-9).
We report the probability of identification within rank 1 averaged over all the pairs (with standard deviation).
We report results for both electricity and electricity and gas consumption features using the best model in each setting (CNN-LSTM and Transformer, respectively).

Table~\ref{tab:freq_repseudonymization} shows that reducing the frequency of re-pseudonymization greatly increases the risk of re-identification, from 75.8\% for a frequency of $F=1$ week to 84.8\% ($F=2$ weeks) and 90.2\% ($F=4$ weeks) for electricity and gas consumption records.
We observe a similar trend for electricity records, as the risk increases from 55.5\% for a frequency of $F=1$ week to 73.6\% ($F=2$ weeks) and 84.6\% ($F=4$ weeks) for electricity consumption records. 

As for more frequent than weekly re-pseudonymization strategies, e.g., on a daily basis, we believe they are unlikely to be used in practice as they would sharply reduce the utility of the data. 
Our method would likely require changes, e.g., to learn what a "normal" Monday looks like compared to a "normal" Sunday, and would probably not work out of the box, i.e., give a good estimate of the risk. However, with changes to the model we believe it may be possible to piece together smart meter records from shorter time periods and we leave the thorough evaluation of the risk for future work.

\textbf{Decreasing training time period.}
Attacks requiring smaller amounts of training data might, in some cases, be more practical to deploy by malicious adversaries. 
Attacks requiring only a month instead of a year of training data would be available to a larger range of attackers who could then use it against re-pseudonymized datasets. 
We investigate the effect of the number of weeks available in the auxiliary dataset $\mathcal{T}_{\text{aux}}$, and consequently of the number of training weeks, on the attack performance. 
More specifically, we retrain each model on the last $M'$ weeks of training data available in the dataset: $M-M'+1, \ldots, M$, varying the number of training weeks $M'$ from 4 to 40.

\begin{figure*}[htbp!]
\centering

\subfigure{%
\includegraphics[width=0.3\textwidth]{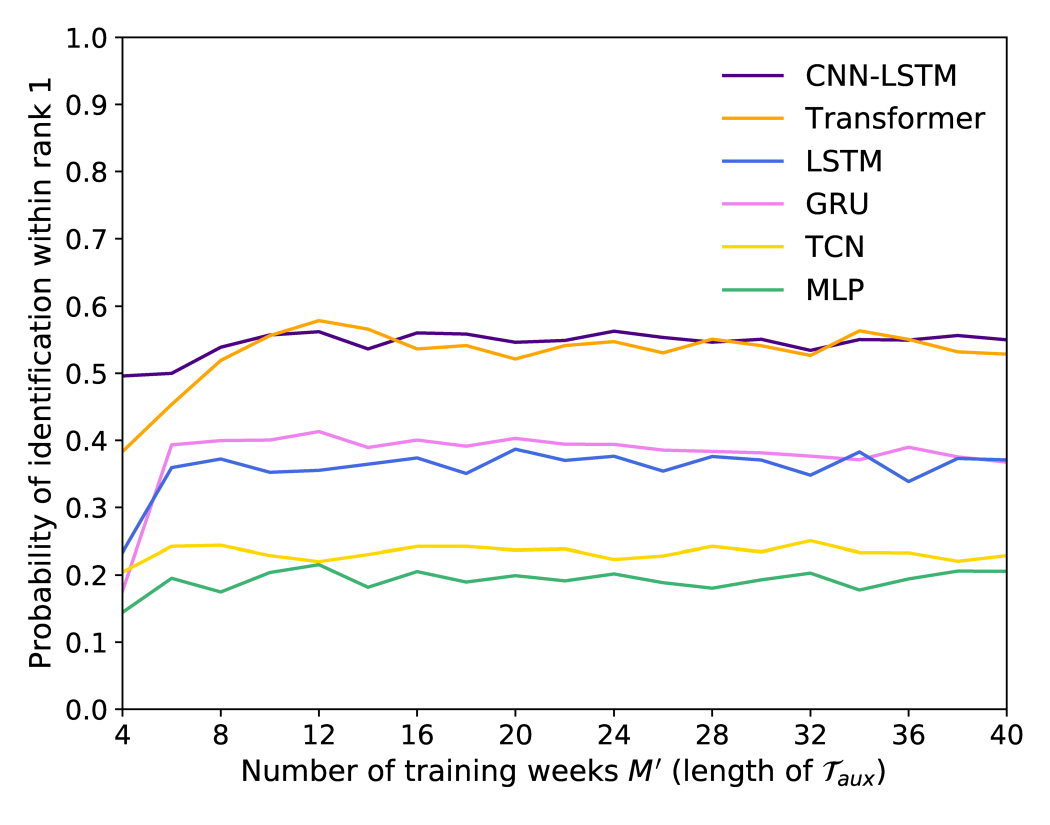}%
}\hfil
\subfigure{%
\includegraphics[width=0.3\textwidth]{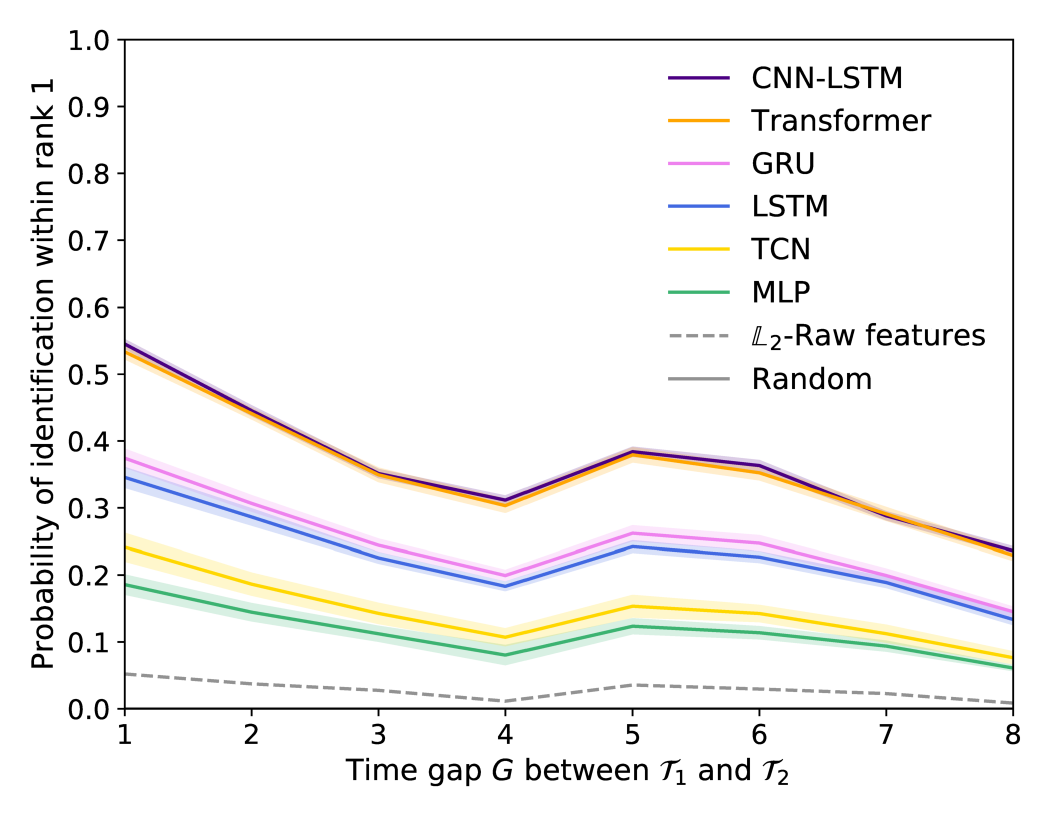}%
}\hfil
\subfigure{%
\includegraphics[width=0.29\textwidth]{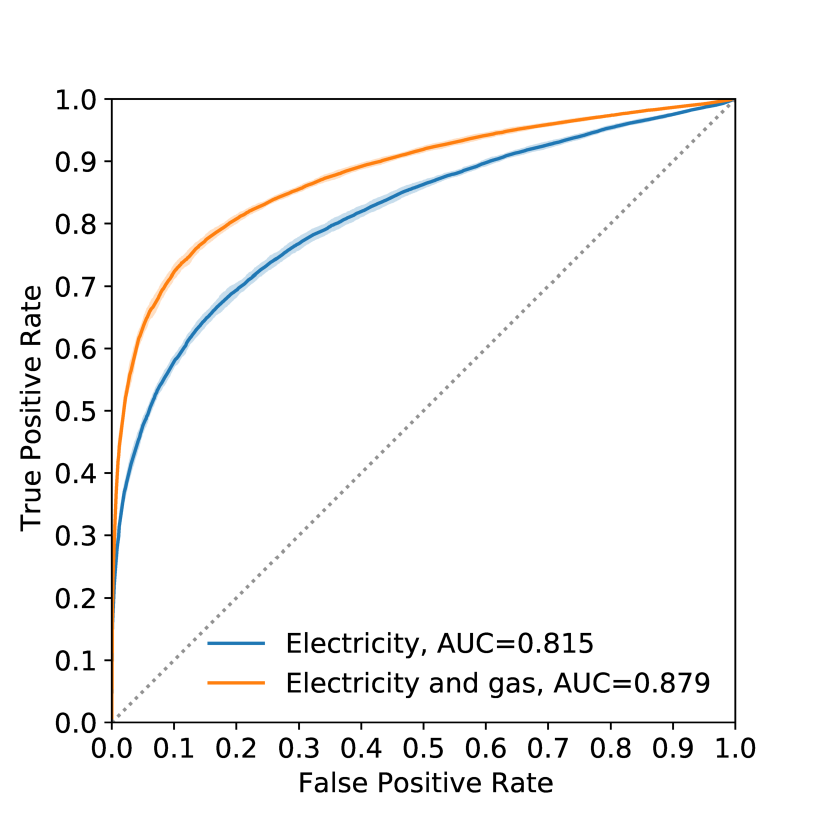}%
}

\caption{Attack performance when varying the number of training weeks of the embedding model (left) and varying the time gap between the re-pseudonymized weeks $\mathcal{T}_1$ and $\mathcal{T}_2$ (middle, mean with std over 10 runs). We further show the ROC curve of a meta-classifier aiming to infer whether the top-1 match is correct based on the gap statistics (right, mean with std over 10 runs).}
\label{figure:secondary_results}
\end{figure*}

We use the first $M'-2$ weeks to generate triplet samples, and the last two weeks as validation data for early stopping, except for $M'=4$.
In this case, as there are fewer data available, we use the first three weeks to generate triplet samples, and the third and fourth weeks as validation data for early stopping.  
As the number of hyperparameters is very large, we used for simplicity the best hyperparameters found when training on $M'=M=40$ weeks. 
Fig.~\ref{figure:secondary_results} (left) shows that our embedding-based approach is sample-efficient on top of being very accurate.

\textbf{Increasing time gap.}
We analyze how the time gap between the re-pseudonymized weeks $\mathcal{T}_1$ and $\mathcal{T}_2$ influences the attack performance. 
One could think that user profiles remain stable over time, as people have daily routines that keep their consumption behavior constant. 
Households may however use more electricity during cold or hot days or may leave home for holidays, which can drastically change their consumption pattern. 
Additionally, depending on the occupation of the household members, e.g., students, there can be a high variability in consumption from one week to another, regardless of the time of the year.
We investigate the impact of the time gap on the attack performance.
We apply our models (trained on weeks 1 to $M$) to the re-pseudonymized weeks $M+1$ ($\mathcal{T}_1$) and $M+1+G$ ($\mathcal{T}_2$) under scenario (I): $P_1=P_{\text{aux}}$, varying G between 1 and 8 weeks.
Fig.~\ref{figure:secondary_results} (middle) shows that the performance of our models decreases overall as the time gap increases.
We also observe a steep drop after $G=4$ weeks, which could be due to an external event leading to changes in the consumption behavior of users.

\begin{table}[t!]
\centering
    \caption{Attack performance (mean with std) for different re-pseudonymization frequencies $F=|\mathcal{T}_1|=|\mathcal{T}_2|$. 
    }
    \label{tab:freq_repseudonymization}
    \begin{tabular}{ccc}
    \toprule
     \textbf{$F=|\mathcal{T}_1|=|\mathcal{T}_2|$} & \textbf{Electricity} &
     \textbf{Electricity and gas}    
     \\ 
     \midrule
     1 week & 55.5 (4.9) & 75.8 (3.8)
     \\
     2 weeks & 73.6 (1.7) & 84.8 (3.7) 
      \\ 
     4 weeks & 84.2 (0.9) & 90.2 (0.8) 
     \\ 
     \bottomrule
    \end{tabular}
\end{table}

\begin{table}[t!]
\centering
    \caption{Attack performance (mean with std over 10 runs) when lowering the precision of smart meter records. 
    }
    \label{tab:rounding_precision}
    \begin{tabular}{ccccc}
    \toprule
     \textbf{Features} & \multicolumn{4}{c}{\textbf{Number of significant digits $n$}}  \\
     & No rounding & $n=3$ & $n=2$ & $n=1$\\
     \midrule
     Elec. & 54.5 (0.6) &  54.5 (0.6) & 53.8 (0.6) & 30.0 (1.1) \\
     \midrule
     Elec. and gas & 73.4 (0.9)  & 73.4 (0.9) & 73.4 (1.0) & 66.9 (0.8) \\
     \bottomrule
    \end{tabular}
\end{table}

\textbf{Rounding precision.}
We study the impact of the rounding precision of the smart meter records on the attack performance.
We round the weekly re-pseudonymized smart meter records in weeks $\mathcal{T}_1$ and $\mathcal{T}_2$ to $n$ significant digits for $n \in \{1, 2, 3\}$, e.g., for $n=1$ a reading of 0.2567 kWh is rounded to 0.3.
Then, we re-run the attack using our best model for electricity and electricity and gas consumption records (CNN-LSTM and Transformer, resp.).
The models are trained on unrounded smart meter records of $D_\text{aux}$.

Table~\ref{tab:rounding_precision} shows that the attack performance is not impacted by the rounding precision, except for $n=1$ where it drops from 54.5\% (no rounding) to 30.0\% ($n=1$) for electricity records and from $73.4\%$ (no rounding) to 66.9\% ($n=1$) for electricity and gas consumption records. 
However, rounding to $n=1$ significant digits impacts utility, as, e.g., 23\%, 36\%, and 25\% of the electricity consumption values are rounded to 0, 0.1, and 0.2 respectively. 
This makes data unusable, e.g., for an energy reduction program where reductions of a few percentage points becomes invisible.

\textbf{Inferring whether the top-1 match is correct.} While our attack correctly identifies a large fraction of users: 54.5\% and 73.4\% in electricity (E) and electricity and gas (E+G) consumption records, the attacker does not know  whether the top-1 candidate is correct.
We here show that the attacker can distinguish between correct and incorrect top-1 candidates based on the ``gap statistic''~\cite{paskov2010regularization,narayanan2012feasibility}, i.e., the difference between the distances of the second and first candidate, $d\big( E(r^{p_2}_{\mathcal{T}_1}), E(r^{p_0}_{\mathcal{T}_2})\big)-d\big( E(r^{p_1}_{\mathcal{T}_1}), E(r^{p_0}_{\mathcal{T}_2})\big)$.
We compute the Receiver Operator Characteristic (ROC) curve of a meta-classifier that uses the gap statistic to infer whether the top-1 candidate returned by the attack is correct.
We report the True Positive Rate for all False Positive Rates between 0 and 1, with a step of 0.001, averaged over 10 model seeds together with the standard deviation.
The results we report are computed on the profiling attack results obtained on weeks $M+1$ ($\mathcal{T}_1$) and $M+2$ ($\mathcal{T}_2$).
Fig.~\ref{figure:secondary_results} (right) shows that the meta-classifier 
obtains a high Area Under the Curve (AUC) of 0.815 and 0.879 on E and E+G records, respectively.
Our attack achieves, e.g., for E+G records and a low false positive rate of 5\%, a true positive rate of 63.3\%, meaning that 0.633*73.4=39.8\% of users would be identified with 95\% confidence.

\textbf{Relevance of hourly granularity.}
Many examples of smart meter record-based sensitive inferences cited in the introduction are derived from more frequent readings than the hourly data available in our dataset.
However, our attack can be trivially applied to finer-grained data by simply aggregating the available data at the hourly level.
Furthermore, extending the attack to operate on such finer-grained records is likely to lead to better attack accuracy, due to more detailed information being available on each household.
This means that the risk posed by profiling attacks to finer-grained records (that are more likely to lead to highly sensitive inferences), is likely to be even higher than the one reported in this work.

\textbf{Limitations.} Our attack succeeds very well even in large populations, e.g., 54.5\% of the time against 5k households, while one might think the risk is 1 in 5k.
This strongly suggests that weekly re-pseudonymized smart meter data is unlikely to be considered anonymous according to the EU's GDPR.
However, our results also suggest that it might not always be possible to piece together enough information to recompose a month and thus lead to successfully re-identify a natural person based on aggregate monthly consumption~\cite{cleemput2018pseudonymization}.
We show bi-weekly re-pseudonymization to be much easier to break, suggesting that it should not be used if re-identification based on monthly aggregate consumption is deemed likely by the data controller.
Finally, new combined profiling and matching attacks could be developed to, e.g., average from 3 weeks to get the approximate monthly consumption.

%% file: sections/conclusion.tex
\section{Conclusion}
\label{sec:conclusion}

Smart meters are currently being deployed at a fast rate throughout the world. They are believed to be a key element in reducing energy consumption and enabling more efficient management of resources. 
Pseudonymization and frequent re-pseudonymization are typically used to preserve the privacy of individuals in these datasets.
In this paper, we use state-of-the-art deep learning techniques to show how users are identifiable at scale and with limited auxiliary information, raising questions on the efficacy of even weekly re-pseudonymization techniques. 
We further show how the identification performance decreases as the size of the population grows and that information about other utility consumption and decreasing the frequency of re-pseudonymization further increase the risk.
Our results strongly suggest that even frequent re-pseudonymization is not an effective risk reduction method.
Instead, we believe privacy engineering techniques such as query-based systems will allow smart meter data to be used while preserving the privacy of individuals.

%% file: sections/appendix.tex
\section{Appendix}
\subsection{Ethics consideration}\label{appendix:ethics_consideration}

In this paper, we analyze the effectiveness of weekly re-pseudony-mization against deep learning-based profiling attacks on smart meter data. We hope our work showing the ineffectiveness of re-pseudonymization risk mitigation strategies to better inform their use, e.g., by encouraging practitioners to implement security measures such as access controls and query-based systems. 

We are aware that our results might increase the risk of deep learning-based profiling attacks being used in practice. We however believe that, given the scope and reach of smart meter data collection, the benefits of our findings being disseminated to the research community, practitioners in the field, and the general public, strongly outweigh the risks. To further limit the risk, we will only make our code available to other researchers upon request and will delete the models upon publication.

\begin{figure}[t!]
    \centering
    \includegraphics[width=0.9\linewidth]{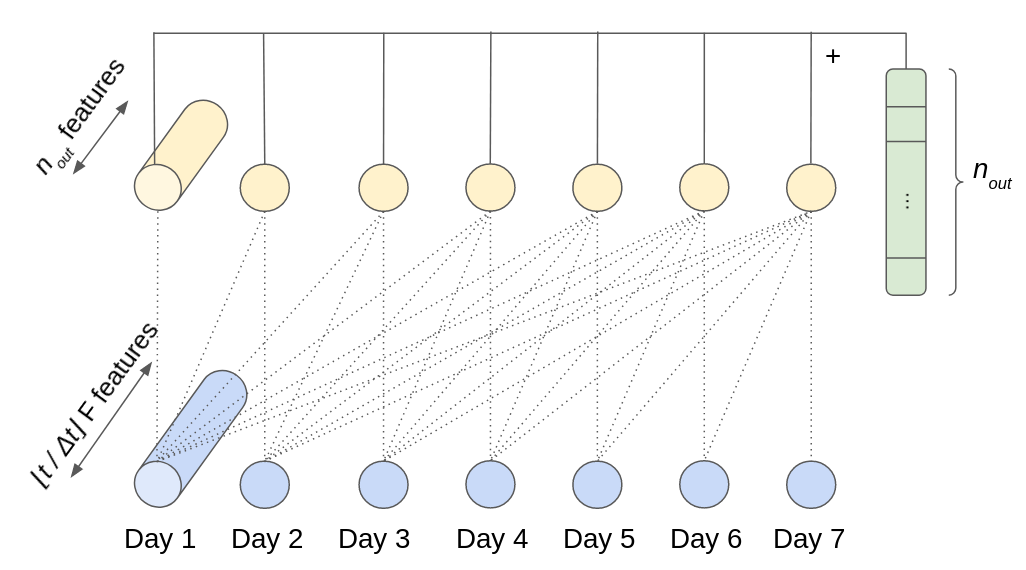}
     \caption{\textbf{Using a TCN for feature extraction.}}
    \label{fig:tcn_input}
\end{figure}

\subsection{Illustration of neural network architectures}
Fig.
~\ref{fig:tcn_input} illustrates the 
TCN architecture used in this paper.
It consists of one temporal block layer that uses a kernel size equal to the number of days in the sequence. 
For clarity, our illustration does not show the padding and we refer the reader to the original paper for complete details~\cite{bai2018empirical}. 
In this example, the embedding is computed at weekly level, i.e., the sequence contains features from $D=7$ consecutive days.

\subsection{Triplet loss details}
\label{appendix:triplet_loss}
We provide additional details about the triplet loss.
It is computed over an \textit{anchor sample} and a \textit{positive sample} coming for the same user $p$ but different time periods and a \textit{negative sample} coming from a different user $p' \neq p$ and any time period. 
The triplet loss is positive (and its gradient is non-zero) in two scenarios: (1) when the distance between the anchor and the positive sample is larger than the distance between the anchor and the negative sample, indicating that the embeddings of $p$ are not well differentiated from those of $p'$ and (2) when the distance between the anchor and the positive samples is lower than the distance between the anchor and the negative samples, but only by a quantity smaller than the margin $\lambda$.
The weights will thus be updated so as to push the embeddings of the anchor and negative samples further away from each other.

\subsection{Training details}
\label{appendix:training_details}
\textbf{General approach.} Training our embedding model (both the optimization and the neural network architecture) requires selecting appropriate hyperparameters that obtain the best result.
There is no theory to support what the best values are, a priori, on a given dataset.
Thus, we performed hyperparameter tuning on the validation data to select the best values. 
We first tried out the choices made in related works (e.g., $\mathbb{L}_2$ normalization is standard with the triplet loss~\cite{schroff2015facenet,crectu2022interaction}) and then explored different choices for these parameters to make sure we do not underfit our models. 
More specifically, we ran preliminary experiments to find plausible ranges for each hyperparameter, using the performance on the validation weeks to guide the hyperparameter selection. 
Then, we performed an extensive hyperparameter search to find the best choice. 
This approach is standard with practitioners when developing a method having too many hyperparameters to perform a grid search over all the possible combinations. 

\textbf{Training details.} The neural network parameters are optimized using mini-batch gradient descent with a batch size of $B=64$, the AdamW optimizer~\cite{loshchilov2017decoupled} and $\mathbb{L}_2$ regularization. We use a margin parameter of $\lambda=1$ for the triplet loss. We optimize each model for a maximum of 300 epochs, using early stopping with a patience of 10 epochs and gradual lowering of the learning rate. Specifically, if the probability of identification within rank 1 computed over the validation weeks does not improve for 10 consecutive epochs, we divide the learning rate by 2, and stop the training if the learning rate becomes smaller than $10^{-5}$.

\textbf{Hyperparameter search.} 
For each architecture, we perform a grid hyperparameter search over the learning rate $\eta \in \{ 0.001, 0.005 \}$, the weight decay parameter $\alpha \in \{0.01, 0.005\}$, the number of architecture layers $L$ (when applicable), and the window lag parameter $l \in \{4, 5, 6, 7\}$. 
We found that using triplet samples from overlapping disjoint weeks ($l \leq 6$) performs better than using triplet samples from disjoint weeks ($l=7$) for some architectures. 
On the one hand, the former setup could make the learning task easier, as the embeddings are -- by design -- more similar when compared to the latter setup in which they are computed from disjoint periods. 
On the other hand, as the lag decreases, the samples become more dependent, meaning that the model is more likely to underperform at test time when applied to disjoint weeks.
As the optimal window lag parameter could depend on the model, we found it important to tune it. 
The best model is always selected based on the performance on the validation weeks.

\subsection{Dataset preprocessing}\label{appendix:data_preprocessing}
We here provide justification for our data preprocessing choices.
Like any empirical attack, our results are only a lower bound for the privacy risk, and it is possible (although unlikely) that other preprocessing decisions further improve performance. 
Both (1) replacing multiple recordings of the same timestamp with a single value and (2) filling the consumption values for missing timestamps were necessary preprocessing steps.
This is because neural networks expect one (and only one) value for each feature (here, a timestamp). 
For (1), it was not clear a priori what the best mapping is, and we found taking the average to be a good rule of thumb. Taking the median was an alternative but we do not believe it would have significantly impacted the results. 
For (2), we experimented with zero- and average value-filling, and found the first method to lead to better results.
Outlier clipping (3) is not necessary for running the attack.
We performed it because we scale features by the mean and standard deviation over all users in the training dataset before feeding them to the embedding model, and the mean is sensitive to outlier values.
Note that we only clipped outliers in the gas dataset.

\subsection{Literature baselines}
\label{appendix:literature_baselines}
\textbf{Buchmann~\cite{buchmann2013re}.} This approach represents the weekly consumption behavior of a user via 12 hand-engineered features such as the overall consumption or the average wake-up hour. 
The method performs similarity-based matching between the feature vectors of a target user from a week and the feature vectors of all the users from the previous week. 
The similarity between two feature vectors is quantified as a linear combination of either absolute or relative measures of difference between feature values. 
Given two users $p, q \in P$ with respective records $R^p_{\mathcal{T}}$ and $R^q_{\mathcal{T}}$ in week $\mathcal{T}$ and features equal to $E_{\text{B}}(R^p_{\mathcal{T}}), E_{\text{B}}(R^q_{\mathcal{T}}) \in \mathbb{R}^{12}$, the absolute difference between the values for the $f$-th feature is equal to $|E_{\text{B}}(R^p_{\mathcal{T}})_f - E_{\text{B}}(R^q_{\mathcal{T}})_f|$ and the relative difference is equal to $|2 (E_{\text{B}}(R^p_{\mathcal{T}})_f - E_{\text{B}}(R^q_{\mathcal{T}})_f)/ (E_{\text{B}}(R^p_{\mathcal{T}})_f - E_{\text{B}}(R^q_{\mathcal{T}})_f)|$, $f=1, \ldots, 12$.
The differences for each feature are mapped to 0 if they are lower than the 90\% quantile of observed differences between the values for that feature and the same user in two different weeks. 
Otherwise, the differences are scaled by subtracting this quantile and dividing by the standard deviation of the lowest 90\% differences. 
In the original work, the quantile and standard deviation are estimated from differences computed on a user set disjoint from the test user set and on their data from the test weeks. 
For a fair comparison with our setup, we estimate the two quantities using data from all pairs of consecutive weeks in the auxiliary dataset: $(I, I+1), 1 \leq I \leq M-1$. 
To optimize the weights of the linear combination, the original paper develops a static and two linear programming approaches. 
In line with Tudor et al.~\cite{tudor2013analysis}, we implement the static approach assigning a weight value of 1. 
Finally, while the original paper reported superior performance of relative differences compared to the absolute differences, we found that using both yields better performance.

\textbf{Tudor~\cite{tudor2015study}.} This approach simplifies the approach of Buchmann et al~\cite{buchmann2013re} by removing the quantile and standard deviation estimation procedure, and therefore the need of ground truth pairs from different weeks. The method computes a smaller set of 5 features $E_{\text{T}} : \mathbb{R}^{168\times F} \longrightarrow \mathbb{R}^5$ and performs Euclidean distance-based matching. In the original paper, when an identity is assigned to a target user, it is also removed from the set of candidates so that the identity cannot be assigned to the (yet) unmatched target users. To be able to report the probability of identification within rank $R$ for $R>1$, we run the attack independently against each target user.

\textbf{Jawurek~\cite{jawurek2011smart}.} This approach trains $N$ one-vs.-rest support vector machine (SVM) classifiers to separate the smart meter records of different users. 
The daily records are represented as binary feature vectors of size $24 \times b$, where each feature corresponds to a consumption range (among $b=100$ ranges) and an hour in the day. 
For efficiency reasons, as we have $M=40$ weeks of training data available and thousands of classes $N>5$k, we use weekly records to train the classifier and test on data from the target week. 
The features in a week are obtained by summing the binary feature vectors from each day $E_{J}: \mathbb{R}^{168\times F} \longrightarrow \mathbb{R}^{2400\times F}$. 
We use the default implementation of support vector classification from the Scikit-learn 0.24.2 library.

\textbf{Faisal~\cite{faisal2015quantity}.} This approach trains a $N$-way decision tree classifier on daily smart meter recordings. We train the classifier on all the users' daily samples available in the first $M$ weeks, totaling $7\times N \times M$ samples. 
To classify data from the target week at test time, we average out the classifier's scores over the seven days and select the highest scoring class. We implemented both a decision tree and a random forest (an ensemble of decision trees) and found the latter to perform better in practice. 
We use the random forest implementation from Scikit-learn 0.24.2 and 100 trees. 
We set the minimum fraction of samples required for each split to 0.005.